%% file: Combined.tex
\documentclass[reprint,aps,prb,graphicx,superscriptaddress]{revtex4-2}

\usepackage{graphicx}
\usepackage{amsmath} 	
\usepackage{xcolor}
\usepackage[utf8]{inputenc} 
\usepackage[T1]{fontenc}
\usepackage[rightcaption]{sidecap}
\usepackage{upgreek}
\usepackage{float}

\let\oldAA\AA
\renewcommand{\AA}{\text{\normalfont\oldAA}}

\newcommand{\Table}[1]{Table~\ref{#1}}

\newcommand{\Fig}[1]{Fig.~\ref{#1}}

\newcommand{\Ef}{E_{\textrm{F}}}
\usepackage{xspace}
\DeclareRobustCommand{\moire}{moir\'{e}\xspace}

\begin{document}
\title{Optical switching of ferro-rotational charge-density wave states}

\author{Wayne~Cheng-Wei~Huang}
\affiliation{Department of Ultrafast Dynamics, Max Planck Institute for Multidisciplinary Sciences, 37077 G\"{o}ttingen, Germany}
\affiliation{IV. Physical Institute--Solids and Nanostructures, University of G\"{o}ttingen, 37077 G\"{o}ttingen, Germany}
\affiliation{Institute of Physics, Academia Sinica, Taipei 115201, Taiwan (R.O.C.)}

\author{Sai~Mu}
\affiliation{Center for Experimental Nanoscale Physics, Department of Physics and Astronomy, University of South Carolina, Columbia, South Carolina 29208, USA}

\author{Gevin~von~Witte}
\affiliation{Department of Ultrafast Dynamics, Max Planck Institute for Multidisciplinary Sciences, 37077 G\"{o}ttingen, Germany}
\affiliation{Institute for Biomedical Engineering, University and ETH Zurich, Zurich 8092, Switzerland}
\affiliation{Institute of Molecular Physical Science, ETH Zurich, Zurich 8093, Switzerland}

\author{Yanshuo~Sophie~Li}
\affiliation{IV. Physical Institute--Solids and Nanostructures, University of G\"{o}ttingen, 37077 G\"{o}ttingen, Germany}

\author{Felix~Kurtz}
\affiliation{Department of Ultrafast Dynamics, Max Planck Institute for Multidisciplinary Sciences, 37077 G\"{o}ttingen, Germany}

\author{Sheng-Hsiung~Hung}
\affiliation{Department of Physics, National Tsing Hua University, Hsinchu 30013, Taiwan (R.O.C.)}

\author{Horng-Tay~Jeng}
\affiliation{Department of Physics, National Tsing Hua University, Hsinchu 30013, Taiwan (R.O.C.)}
\affiliation{Institute of Physics, Academia Sinica, Taipei 115201, Taiwan (R.O.C.)}
\affiliation{Physics Division, National Center for Theoretical Sciences, Taipei 10617, Taiwan (R.O.C.)}

\author{Kai~Rossnagel}
\affiliation{Institute of Experimental and Applied Physics, Kiel University, 24098 Kiel, Germany}
\affiliation{Ruprecht Haensel Laboratory, Deutsches Elektronen-Synchrotron DESY, 22607 Hamburg, Germany}

\author{Jan~Gerrit~Horstmann} 
\affiliation{Department of Ultrafast Dynamics, Max Planck Institute for Multidisciplinary Sciences, 37077 G\"{o}ttingen, Germany}
\affiliation{Department of Materials, ETH Zurich, 8093 Zurich, Switzerland}

\author{Claus~Ropers}
\email{Email: claus.ropers@mpinat.mpg.de} 
\affiliation{Department of Ultrafast Dynamics, Max Planck Institute for Multidisciplinary Sciences, 37077 G\"{o}ttingen, Germany}
\affiliation{IV. Physical Institute--Solids and Nanostructures, University of G\"{o}ttingen, 37077 G\"{o}ttingen, Germany}

\begin{abstract}
Tailored optical excitations can steer a system along non-equilibrium pathways to metastable states with specific structural or electronic properties. The light-induced hidden state of 1T-TaS$_{2}$, with its strongly enhanced conductivity and exceptionally long lifetime, represents a unique model system for studying the ultrafast switching of correlated electronic states. We use surface-sensitive electron diffraction in combination with a femtosecond optical quench to reveal the coexistence of both charge-density-wave (CDW) 2D chiralities as a structural characteristic of the hidden state, corresponding to coexisting ferro-rotational CDW states. Density functional theory (DFT) simulations of interfaces between opposite CDW 2D chiralities predict a higher-level, fractal-type moir'{e} superstructure with a kagome band structure near the Fermi energy. More broadly, these findings suggest that heterochiral interfaces in CDW systems provide an additional structural degree of freedom, expanding the possibilities for electronic control via twist-angle engineering.
\end{abstract}

\maketitle

\textbf{ }

Ultrafast control of functional materials via non-equilibrium pathways promises access to tailored electronic and structural properties \cite{Bao2022, Torre2021, Basov2017,Stojchevska2014, Horstmann2020, Rini2007}. A particularly prominent example is the 1T-TaS$_{2}$ hidden state \cite{Stojchevska2014}, which may enable novel non-volatile all-electronic memory devices \cite{Vaskivskyi2016,Burri2025}. The long-lived metastable hidden state exhibits metallic conductivity \cite{Stojchevska2014,Gao2022,Svetin2017} and is generated from the insulating low-temperature commensurate charge-density wave phase (C-phase) \cite{vonWitte2019} by applying single laser or voltage pulses \cite{Stojchevska2014, Gerasimenko2019, Cho2016, Ma2016, Vaskivskyi2015}. Structural changes associated with this state have been probed by scanning tunneling microscopy (STM) and X-ray diffraction, yielding an emergent mosaic network of domain walls and the breakdown of interlayer dimerization \cite{Gerasimenko2019, Cho2016, Ma2016, Stahl2020}. However, due to the large variability in generation conditions and properties associated with the hidden state, further systematic structural studies of metastable and mixed-order states in this material are needed, including thermally quenched pathways that stabilize a mixed-CDW regime at equilibrium \cite{Delatorre2025}.

In the present work, we reveal a new structural characteristic concurring with the hidden state: coexistent charge-density-wave (CDW) 2D-chiralities that represent opposite domain states of a ferro-rotational order parameter defined by the in-plane orientation of the CDW superlattice. Using low-energy electron diffraction (LEED) \cite{Vogelgesang2018} following a single-pulse optical quench \cite{Stojchevska2014, Vaskivskyi2015, Gao2022, Gerasimenko2019, Stahl2020}, our experiments demonstrate the emergence of CDW domains of opposite 2D-chirality with respect to the initial state. The optically prepared state represents a multidomain configuration with imbalanced populations of two ferro-rotational domains.
The electronic origin of this structural transformation is verified by double-pulse excitation \cite{Maklar2023, Ravnik2018}. Fluence-dependent diffraction intensities of the newly formed domains of opposite 2D-chirality show a threshold behavior for 13 measured samples. The measured threshold fluence varies among the ensemble of crystals and is in general accordance with previously observed conditions for hidden-state generation \cite{Stojchevska2014, Gerasimenko2019, Stahl2020}. At high fluence, most specimens exhibit a saturated diffraction intensity of the minority 2D-chirality between 2\,\% and 7\,\% of the majority 2D-chirality, while two samples reach about 10\,\% without a clear sign of saturation. Our study implies extended CDW moir\'{e} interfaces present in the laser-induced state, for which DFT simulations predict a new type of kagome charge order that supports in-plane metallic conductivity. Alongside disordered CDW stacking arrangements \cite{Cho2016, Ma2016}, such 2D-heterochiral interfaces may contribute to the electronic properties of the hidden state, and more generally, suggest active laser control of interface 2D-chirality also in other contexts \cite{Ohta2021, Stojchevska2018, Song2022, Liu2023}.

\begin{figure*}[!ht]
\centering
\includegraphics[width = \textwidth]{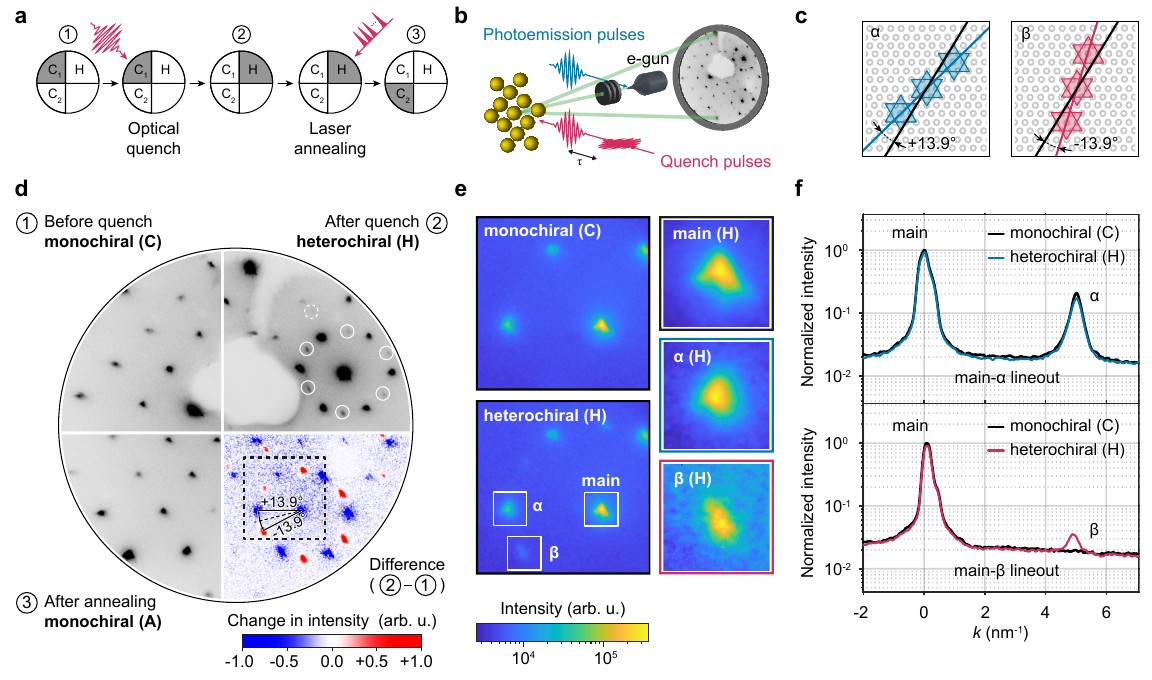}
\caption{\textbf{LEED images of the laser-induced 2D-heterochiral CDW state.} 
\textbf{a}, Sketch of the optical excitation sequence, with quadrants representing the LEED images in \textbf{d}.
\textbf{b}, Experimental setup for single- and double-pulse quenches monitored by LEED.
\textbf{c}, The two 2D-chiralities of the 1T-TaS$_{2}$ CDW superstructure, rotated by $\pm13.9^{\circ}$ relative to the host lattice (black line).
\textbf{d}, Sections of LEED images in the sequence from the C-phase to the 2D-heterochiral CDW state and back to a 2D-monochiral state. A single-pulse quench of the C-phase (top-left panel, sample \#1, pulse fluence: 1.7\,mJ\,cm$^{\textrm{-}2}$) generates first- and second-order $\beta$ superstructure peaks (solid and dashed circles, respectively, in the top-right panel). Annealing with a laser pulse train reverses the sample to a 2D-monochiral state (bottom-left panel) indistinguishable from the initial C-phase. A difference image (bottom-right panel) highlights the emergence of $\beta$ peaks and the suppression of $\alpha$ peaks in the 2D-heterochiral state.
\textbf{e}, Close-up LEED images of the C-phase and the 2D-heterochiral state. Left column: area denoted by the black dashed rectangle in \textbf{d}. Right column: close-ups of main lattice and superstructure peaks (white solid rectangles in the bottom-left panel), normalized to individual peak heights. Main lattice peaks stem from the undistorted lattice, whereas $\alpha$ and $\beta$ peaks are associated with the CDW-coupled periodic lattice distortion.
\textbf{f}, Lineouts along main-$\alpha$ and main-$\beta$ directions (bottom-left panel in \textbf{e}) illustrate that the spot profiles of the main lattice and $\alpha$ peaks remain nearly unchanged after the optical quench.}
\label{fig:image}
\end{figure*}

\begin{figure*}[!ht]
\centering
\includegraphics[width = \linewidth]{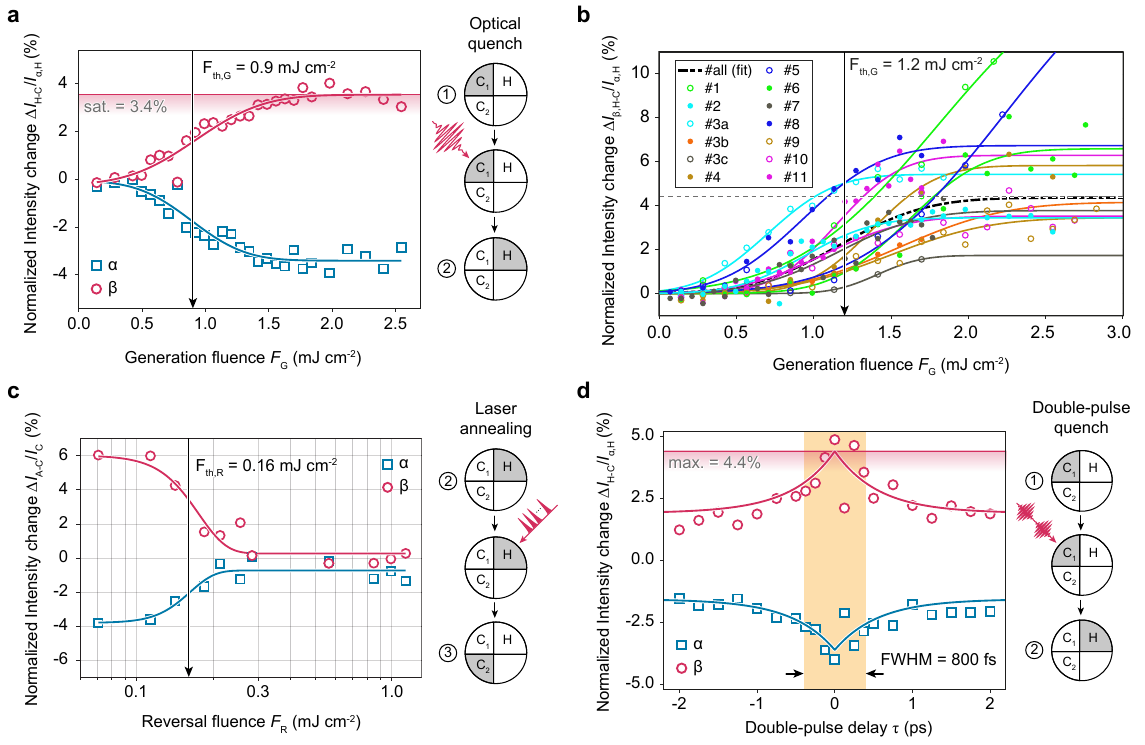}
\caption{\textbf{Generation conditions for the 2D-heterochiral CDW state.} 
\textbf{a}, Normalized intensity changes of $\alpha$ (blue squares) and $\beta$ (red circles) peaks as a function of the single-pulse quench fluence (sample \#2). Both traces exhibit a threshold (black vertical line) and saturation behavior (red horizontal line).
\textbf{b}, Normalized $\beta$ peak intensity after the single-pulse quench. Measurements of the 13 samples are fitted with error functions. The obtained threshold fluences vary from 0.7\,mJ\,cm$^{\textrm{-}2}$ (\#3a) to 2.2\,mJ\,cm$^{\textrm{-}2}$ (\#5), and saturation levels range from 2\,\% (\#3c) to about 7\,\% (\#6), with two samples (\#1 and \#5) showing no signs of saturation. Considering all data, an average threshold fluence of 1.2\,mJ\,cm$^{\textrm{-}2}$ and saturation intensity ratio $I_{\beta}/I_{\alpha} \approx 4\,\%$ are determined. 
\textbf{c}, Normalized intensity changes as a function of the annealing pulse fluence (sample \#3a). The $\alpha$ trace does not fully return to 0\,\% for annealing fluences up to 1\,mJ\,cm$^{\textrm{-}2}$, possibly due to some irreversible changes of the sample. 
\textbf{d}, Double-pulse optical quench (sample \#4). Normalized intensity changes as a function of the pulse separation $\tau$ show an enhancement for $|\tau| \le 400\,\textrm{fs}$ (shaded area). At a combined pulse fluence of 1.14\,mJ\,cm$^{\textrm{-}2}$, the $\beta$ peak intensity generated by the double-pulse quench varies between 4.9\,\% ($\tau = 0\,\textrm{ps}$) and 1.2\,\% ($\tau = -2\,\textrm{ps}$). (All data obtained by averaging the intensities of 15 CDW superstructure peaks in each LEED image. Data sets \#3a, \#3b, \#3c taken on the same sample but with slightly different surface conditions and laser pulse durations.)}
\label{fig:curve}
\end{figure*}


\section*{Experimental results}

We probe the structural changes of optically quenched 1T-TaS$_{2}$ crystals with high-coherence microbeam LEED \cite{Vogelgesang2018}. Our most important observation is the emergence of additional CDW diffraction peaks with a 2D-chirality opposite to that of the initial state (\Fig{fig:image}d, sample \#1). Specifically, we find new CDW superstructure peaks at the opposite rotation angle with respect to the main lattice after the sample is quenched by a single laser pulse. In the absence of further perturbations, this new structural feature persists beyond several hours. We denote the CDW superstructure peaks in the initial C-phase as $\alpha$ and the emergent peaks as $\beta$. These peaks correspond to the two ferro-rotational domain states of the commensurate CDW. The intensity difference of images before and after the optical quench (bottom-right panel of \Fig{fig:image}d) highlights the suppression of the $\alpha$ peaks and the emergence of the $\beta$ peaks in the 2D-heterochiral CDW state. A reversal to the initial LEED pattern, i.e., an erasing of the emergent $\beta$ peaks, can be achieved by illuminating the sample with a femtosecond laser pulse train (repetition rate 100\,kHz, duration of 1\,s) at a fluence of 0.5\,mJ\,cm$^{\textrm{-}2}$. Besides an occasional intensity decrease of the main lattice and $\alpha$ peaks that was observed in some samples, the resulting 2D-monochiral state has a diffraction pattern (bottom-left panel of \Fig{fig:image}d) practically indistinguishable from that of the initial C-phase.  


As shown in \Fig{fig:image}e, the diffraction pattern after the quench reveals the emergence of $\beta$ peaks, whose spot sizes match those of the pre-existing $\alpha$ peaks (\Fig{fig:image}f). The main lattice and $\alpha$ peaks retain their profiles, with only a slight intensity decrease and no noticeable pulse-induced broadening, evidencing the absence of increased structural disorder on a scale below the coherence length of the probe \cite{Storeck2021}. The minimum correlation lengths for both $\alpha$ and $\beta$ 2D-chiralities are approximately 23~nm (see Supplementary Information, Sec. I), suggesting that the $\beta$ domains exhibit structural order beyond several CDW unit cells.

Figure~\ref{fig:curve}a shows the emergence of $\beta$ peaks and a concurrent suppression of the $\alpha$ peak intensity as a function of the single-pulse quench fluence. This measurement is carried out for one of the investigated samples with a randomized sequence (sample \#2), demonstrating that the observations are not affected by sample deterioration. The intensity changes are normalized to the $\alpha$ peak intensity in the 2D-heterochiral state, accounting also for the background. Overall, the intensity changes are on the order of a few percent of the $\alpha$ peak intensity. We find a threshold behavior of the arising $\beta$ peak intensity at a fluence of $F_{\text{th,G}}=0.9$\,mJ\,cm$^{\textrm{-}2}$, which is in close correspondence to the threshold reported for preparing the hidden state \cite{Stojchevska2014, Gerasimenko2019, Stahl2020}. Above the fluence threshold, we observe a saturation in the $\beta$ peak intensity for this sample. Neither additional features in the diffraction pattern nor permanent sample damage are observed for the quench fluence in the saturation regime (no sample damage was observed in LEED images for fluences up to 3.4\,mJ\,cm$^{\textrm{-}2}$). Fluence-dependence measurements of 12 other samples (with a measurement sequence from low to high fluences) show a very similar behavior, albeit with varying threshold and saturation characteristics (\Fig{fig:curve}b). We note that a comparable number of investigated samples did not show the formation of the 2D-heterochiral state, and some only did for a limited time after cleaving. Overall, a very high surface quality, low contamination, and likely low defect density appear to be a prerequisite for generating this state.

By repeatedly preparing the sample in the 2D-heterochiral state using a single-pulse quench at 1\,mJ\,cm$^{\textrm{-}2}$, we find an annealing threshold for the reversal to the 2D-monochiral state for a laser pulse train (repetition rate 100\,kHz, duration 1\,s) at a fluence of $F_{\text{th,R}}=0.16$\,mJ\,cm$^{\textrm{-}2}$ (\Fig{fig:curve}c, sample \#3a). Both the long lifetime and the comparatively low annealing threshold are clear evidence of metastability, as previously reported for the hidden state \cite{Stojchevska2014, Vaskivskyi2015}. The state's high sensitivity to the surface quality also suggests that it resides in a particularly shallow metastable free-energy minimum. The moderately larger width of the CDW peaks compared to the main peaks (see Fig.~S1) implies a shorter correlation length of the superstructure compared to the host lattice, even after this laser reversal to the 2D-monochiral state. This suggests the presence of residual and more stable translational $\alpha$-domains. 

To elucidate the generation mechanism of the 2D-heterochiral state, we employ a double-pulse quench scheme. In \Fig{fig:curve}d, the intensity changes of the $\alpha$ and $\beta$ peaks are measured (with a randomized sequence) as a function of the time delay $\tau$ between two successive quench pulses. The two pulses are orthogonally polarized, and each pulse has a below-threshold fluence of 0.57\,mJ\,cm$^{\textrm{-}2}$ (sample \#4, threshold fluence $F_{\text{th,G}}=1.4$\,mJ\,cm$^{\textrm{-}2}$). An ultrafast enhancement of the arising $\beta$ peak intensity is found at the pulse overlap with an exponential decay time of about 400\,fs, evidencing that switching to the 2D-heterochiral state is enhanced on timescales shorter than electronic relaxation. Although the signal-to-noise ratio of our data does not allow for a determination of possible coherent oscillations, and shorter optical pulses may be necessary to resolve the CDW amplitude mode, the timescale of this enhancement is in good agreement with previous time-resolved measurements of hidden-state generation \cite{Maklar2023, Ravnik2018}. 

\section*{Discussion}

\begin{figure*}[ht]
\centering
\includegraphics[width = \linewidth]{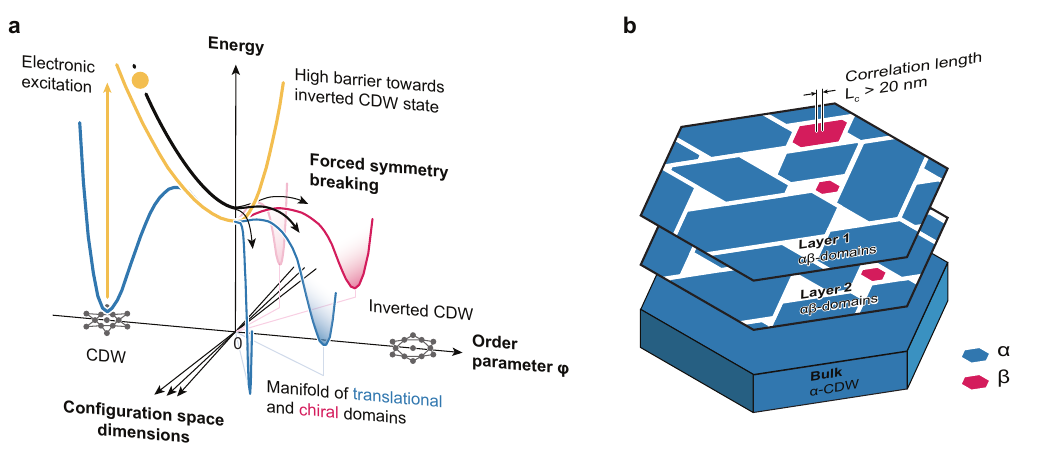}
\caption{\textbf{Proposed generation mechanism and possible structural texture for coexistent CDW 2D-chiralities.} 
\textbf{a}, Suggested mechanism for the generation of a 2D-heterochiral hidden state with translational domains. The illustration depicts simplified free-energy surfaces as a function of the order parameter $\varphi$ of the homogeneous 2D-monochiral C phase, and along other coordinates of the high-dimensional configuration space (schematically indicated as multiple axes). The optically-induced displacive excitation of the CDW causes a rapid evolution of the system towards an inverted CDW state which is energetically highly unfavorable. Collision with this steep barrier leads to a forced symmetry breaking into a distribution of translational and 2D-chiral domains. Concurrent with electronic cooling and structural relaxation, this branching finally results in the formation of a metastable state with quenched disorder. \textbf{b}, From the relative diffraction intensities of $\alpha$ and $\beta$ peaks, the experiments imply a random distribution of both 2D-chiralities in the surface layer. Recent STM measurements corroborate this scenario \cite{Ravnik2023}. Layers in both textures may be covered with translational domains, for which measured diffraction peak widths yield a minimum structural correlation length of about $\geq 20$\,nm. This implies extended interfaces between CDWs of opposite 2D-chirality at or near the surface. 
}
\label{fig:model}
\end{figure*}

\begin{figure*}[ht]
\centering
\includegraphics[width = \textwidth]{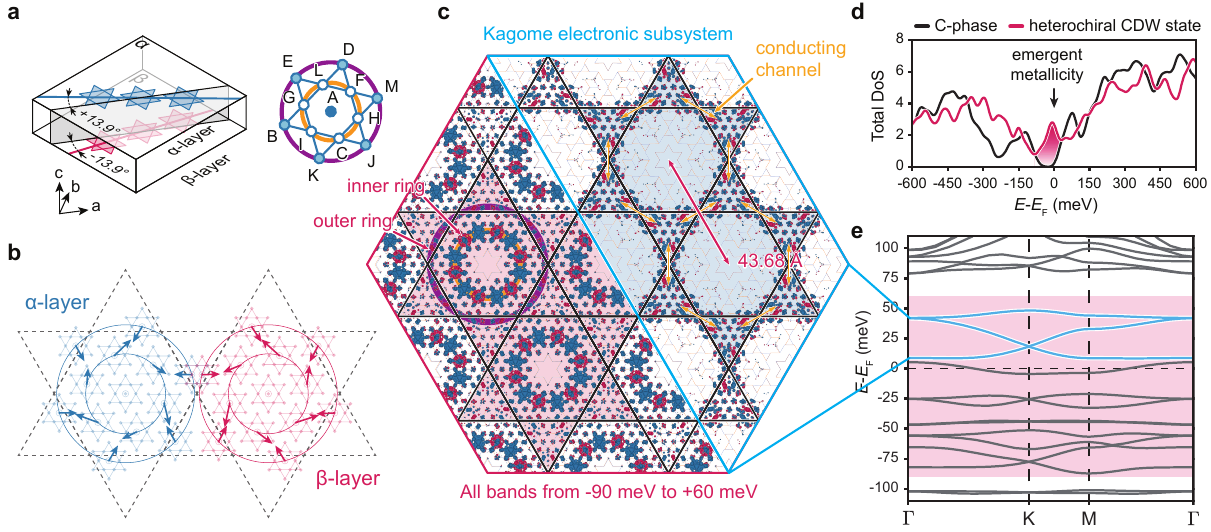}
\caption{\textbf{Emergent kagome system in the CDW moir\'{e} superstructure.}
\textbf{a}, Construction of the CDW moir\'{e} unit cell (see also depiction in Fig.\,S6b). (Left) The structure considered consists of two 1T-TaS$_{2}$ trilayers of opposite CDW 2D-chirality, $\alpha$ (blue) and $\beta$ (red). (Right) The moir\'{e} structure is described by a super-hexagram unit cell containing 13 Ta hexagrams in each layer. Circles denoted by capital letters A to M indicate the center of the Ta hexagrams and the local stacking order of both layers \cite{Lee2019}, with darker filling color representing higher electrostatic energy. The inner and outer rings of the super-hexagram are marked by orange and purple, respectively.    
\textbf{b}, Kagome structural order emerging from the coupling of both layers. A comparison between atomic positions in the CDW moir\'{e} superstructure and the C-phase shows that displacements of the Ta hexagram centers (depicted by arrows) break the translational symmetry of Ta hexagram, resulting in a $13 \times 13$ kagome superlattice. The concentric rings mark the inner and outer rings of the super-hexagrams. The Ta atoms at the center of the super-hexagrams experience no displacement.
\textbf{c}, Energy- and band-resolved electron density distributions. The total electron density for bands from -$90$\,meV to $60$\,meV around $\Ef$ develops a double-ring texture (left, red shading), while the kagome electronic subsystem (right, blue shading) features a conducting network (yellow arrows). Both charge orders follow the same kagome superlattice (black lines) as the atomic structure. Electron density of $\alpha$ and $\beta$ layers are shown in blue and red, respectively. Isosurface values are adjusted for optimal clarity of the charge pattern.
\textbf{d}, Total density of states (DoS) and \textbf{e}, band structure of the 2D-heterochiral CDW state. Several flat bands and a kagome electronic subsystem (highlighted in blue) are visible near $\Ef$.
}
\label{fig:DFT}
\end{figure*}

\subsection{Nature and origin of the 2D-heterochiral CDW state}
The experimental evidence strongly suggests that the observed 2D-heterochiral CDW state is one of the manifestations of the 1T-TaS$_{2}$ hidden state. This view is supported by the close correspondence of experimental conditions to prepare the 2D-heterochiral and hidden states, including the sample temperature, excitation parameters and fluence thresholds, electronic enhancement, surface quality, as well as the long metastable state lifetime and the ability to reverse the state generation.


Switching between different CDW orientations in 1T-TaS\textsubscript{2} has first been observed in the NC-phase at room temperature with the switching involving a transient IC CDW state ($\varphi = 0$\textdegree) \cite{Zong2018,Vogelgesang2018thesis,Delatorre2025}, leading up to \textmu m-sized domains . A transient IC state appears incompatible with the experimental fluences and equilibrium temperatures required for the creation of the hidden state. Therefore, the occurrence of both CDW 2D-chiralities in the hidden states requires a different mechanism, which will be discussed below. Moreover, Ravnik \textit{et al.} \cite{Ravnik2023} observed laser-induced mirrored domains across small areas via STM experiments, which exhibited a structural relaxation within minutes, and which were interpreted as a precursor to the formation of the H-state.

Pump-pump-probe experiments suggest that the hidden state creation depends on the hexagram amplitude mode, eventually involving a transient overshoot into an inverted hexagram state, i.e., the outer twelve Ta atoms are further away from the central Ta atom compared to the high-temperature $1 \times 1$ phase \cite{Maklar2023}. THz spectroscopy and optical reflectivity measurements motivated a Landau-Ginzburg-Wilson theory to model the formation of the H-state after the optical pulse \cite{Gao2022}. In this model, the H-state is described as consisting of domains of the 13 possible stackings (cf. Fig.~\ref{fig:DFT}a) with domain sizes around 10\,nm \cite{Gao2022}. Previous theoretical and experimental works have demonstrated that several of the 13 stackings lead to a metallic behavior \cite{Stahl2020, Ritschel2018, Lee2019, Petocchi2022, Nicholson2022, Salzmann2023, Lee2021, Butler2020, Fei2022}. A hidden state consisting of small domains appears consistent with a Drude-Smith-type conductivity of confined charge carriers as suggested by THz spectroscopy \cite{Gao2022,Smith2001}.
.


Based on the above discussion and results, the formation of the hidden state is qualitatively sketched in \Fig{fig:model}a. First, the electronic excitation by the laser pulse drastically deforms the free-energy potential by strongly populating the amplitude mode of the CDW, initiating an expansion to an inverted CDW state. The expansion from a CDW to an inverted CDW state requires passing close to the high-symmetry state ($\varphi = 0$\textdegree). However, the inverted CDW state is energetically highly unfavorable, and from the high-symmetry state, the system can locally branch off into many of the nearly degenerate configurations, including different stackings and CDW 2D-chiralities (cf. Fig.~\ref{fig:model}b). In this picture, the minority contribution of the $\beta$ CDW orientation, as shown in our LEED images, can be attributed to its somewhat more unfavorable free energy and/or a narrower path across the transition state. The reversal to a 2D-monochiral state can also be understood on the basis of a shallow metastable energy minimum for a $\beta$ domain in an $\alpha$-domain environment. 

This transition bears some resemblance to the Kibble-Zurek mechanism, which describes a quench through a second-order phase transition from a highly symmetric to a spontaneously broken symmetry state \cite{Deutschlander2015, Eggebrecht2017}. In contrast, however, the H-state formation evolves from a symmetry-broken homogeneous phase after excitation into many symmetry-broken local stacking and mirror domains. This likely involves swiftly traversing the high-symmetry state, which motivates denoting the mechanism as a ``forced symmetry breaking'' compared to the spontaneous symmetry-breaking described by Kibble and Zurek.

Dynamical LEED simulations of 1T-TaS\textsubscript{2} \cite{vonWitte2019,Storeck2020} suggest that around 3.7\% of the total signal arises from the subsurface S-Ta-S (tri-)layer (cf. Fig.\,S5, Supplementary Information). The signal of the $\beta$ peaks exceeds 5\% for several samples, and reaches ten percent for two of the samples (cf. Fig.~\ref{fig:curve}b). Therefore, it appears most likely that the $\beta$ domains are randomly distributed in both the surface and subsurface layers.

The rather large variation between the samples remains unclear, especially the two samples not showing signs of a plateau. Assuming that between 5 and 10\% of a given layer is converted to a $\beta$ orientation and the domains being typically larger than 23\,nm, up to 10-20\% of a pair of layers is composed of an $\alpha$-$\beta$ interface. If the metallic conductivity of the hidden state arises from the local conductivities of the different domains, as discussed above based on Ref. \cite{Gao2022}, the $\alpha$-$\beta$ interface might affect the overall hidden state conductivity.
Thus, in the following, we discuss the properties of the $\alpha$-$\beta$ interface based on density functional theory (DFT) simulations.

\subsection*{Properties of $\alpha$-$\beta$ interfaces}

The 27.8$^\circ$ rotational mismatch between CDWs between the $\alpha$ and $\beta$ layers gives rise to a commensurate CDW moir\'{e} superstructure (cf. Fig.\,S6b, Supplementary Information) with an unusually large moir\'{e} length ($\lambda_{m} = 43.68\,\AA$) compared to typical large-angle moir\'{e} materials \cite{Shallcross2010, Tilak2023, Carr2020}. Using DFT simulations, we explore the electronic and structural order of the CDW moir\'{e} interface, considering a simplified atomic structure composed of two 1T-TaS$_{2}$ trilayers of opposite CDW 2D-chirality (\Fig{fig:DFT}a, left). The unit cell of the superstructure contains 13 Ta hexagrams in each trilayer, in a fractal-type arrangement that forms a ``super-hexagram'' (see also Fig.\,S6b, Supplementary Information). The angle between $\alpha$ and $\beta$ super-hexagrams in adjacent layers creates a mixed stacking order \cite{Lee2019} (\Fig{fig:DFT}a, right). 


The main results of our DFT simulations are summarized in \Fig{fig:DFT}b-e. In the CDW moir\'{e} unit cell (cf. Fig.\,S6b, Supplementary Information), the relaxed atomic structure transforms from a ($\sqrt{13}\times\sqrt{13}$)\textit{R}13.9$^{\circ}$ triangular superlattice to a $13\times13$ kagome superlattice (\Fig{fig:DFT}b and Fig.\,S7, Supplementary Information), with the CDW 2D-chirality and Ta hexagrams of each layer remaining unaltered. The atomic displacements are of the order $0.04\,\AA$. They collectively induce an electronic reconstruction, with electron density redistributed into a novel double-ring charge texture (\Fig{fig:DFT}c, left) with a charge-depleted central hexagram. A prominent inner charge ring is composed of 6 hexagrams, while a less pronounced outer ring spans 12 hexagrams shared by adjacent unit cells. This texture may arise from the stacking-dependent electrostatic energy, as discussed in Sec.\,III of the Supplementary Information. Such an electron density distribution may correspond to the ring-like charge structures observed by STM in Refs.~\cite{Ohta2021}~and~\cite{Ravnik2023}, hence favoring the $\alpha\beta$-domain texture (\Fig{fig:model}b, right). The regular $\beta$ domains observed in Ref.~\cite{Ravnik2023} imply regions with a deeper 2D-heterochiral interface. We believe that future spectroscopic STM measurements will elucidate the connection between the observed ring-like charge pattern and the CDW moir\'{e} superstructure introduced here.

Alongside the structural transformation and electronic reconstruction, the density-of-states calculation shows an emergent conducting peak at the Fermi energy ($\Ef$), which strongly contrasts with the electronic gap of the C-phase (\Fig{fig:DFT}d and Fig.\,S9b, Supplementary Information). The band structure calculation further reveals multiple flat bands and Dirac cones emerging near $\Ef$ (\Fig{fig:DFT}e). While the flat bands at $\Ef$ may facilitate correlation effects \cite{Cao2018-1, Cao2018-2}, the lowest conduction bands (within 50\,meV above $\Ef$) feature characteristics of a kagome subsystem \cite{Yin2022}: a Dirac point at K, a van Hove singularity at M and a flat band across $\Gamma$-K-M-$\Gamma$. The corresponding electron density distribution hosts a kagome network of conducting channels (\Fig{fig:DFT}c, right). It is worth noting that the nearly half-filled band at $\Ef$ in \Fig{fig:DFT}d arises naturally from the 2D-heterochiral CDW lattice structure with the electronic reconstruction. 

The DFT simulations indicate that a kagome system emerges from the twisting of the $\sqrt{13} \times \sqrt{13}$R13.9\textdegree\ CDW of 1T-TaS\textsubscript{2} by 27.8\textdegree\ with the hexagrams as large ``virtual'' atoms. The $\sqrt{13} \times \sqrt{13}$ CDW occurs in several transition metal dichalcogenides (TMDCs) and it would be interesting to study if the proposed emergent kagome system is specific to 1T-TaS\textsubscript{2} or rather connected to the basic geometry, i.e., occurring in several related compounds. Ohta \textit{et al.} \cite{Ohta2021} studied an $\alpha$-$\beta$ structure of 1T-TaSe\textsubscript{2} via STM and found a ring structure resembling Fig.~\ref{fig:DFT}c, with a prominent inner ring of six hexagrams and a depleted signal at the center and in the outer ring. These related observations suggest the design of structures connecting the features of moir\'{e} and kagome systems, as also discussed in Ref.~\cite{Yin2022}.


\section*{Conclusions}

Combining femtosecond laser excitation with high-coherence LEED, we observe the coexistence of both CDW orientations  at the surface of 1T-TaS$_{2}$ crystals. Prepared under conditions corresponding to the well-known hidden state in this material, we suggest that the coexistence of both CDW orientations is a structural feature closely linked to the hidden state. Furthermore, we propose that this new structural element arises from a forced symmetry breaking leading to domains with different stackings and CDW orientations. The coexistent ferro-rotational  domains $\geq 20$\,nm in diameter implies the presence of extended $\alpha$-$\beta$ interfaces. Using DFT simulations, we reveal the electronic reconstruction of the double-ring charge texture, identify an emergent kagome system, and obtain a nearly half-filled band at $\Ef$.
At present, it is uncertain to what degree the emergent metallicity of $\alpha$-$\beta$ interfaces contribute to the macroscopic properties of the hidden state, as this likely depends on the actual fraction of interfaces formed. However, future work will allow for large-area fabrication of such CDW moir\'{e} interfaces, enhancing twist-angle electronic engineering by 2D-chirality.

\section*{Methods}

\subsection*{LEED measurements}
In the experiments, large flakes of 1T-TaS$_{2}$ samples (diameter 2-3\,mm, thickness 150\,$\upmu$m) are cleaved under high vacuum and transferred to an ultrahigh vacuum chamber with a base pressure of $2 \times 10^{\textrm{-}10}$\,mbar. The samples are prepared in the C-phase at a temperature of 30\,K using a liquid helium flow cryostat. LEED images in a backscattering geometry near normal incidence (cf. \Fig{fig:image}b) are recorded using a narrow electron beam (150\,$\upmu$m diameter) with high momentum resolution from a millimeter-sized photoelectron gun (sample-to-gun distance 10\,mm, see Ref.~\cite{Vogelgesang2018} for details on the electron source). The low-temperature C-phase of 1T-TaS$_{2}$ exhibits a ($\sqrt{13}\times\sqrt{13}$)\textit{R}13.9$^{\circ}$ CDW superstructure. As our study is concerned with the 2D-chirality of the superstructure \cite{Zong2018, Ishioka2010, Song2022, Liu2023,Delatorre2025}, in each measurement, we begin with a large single-crystalline area of a single CDW 2D-chirality. The diffraction pattern of the initial state (cf. \Fig{fig:image}d, top-left panel) exhibits a hexagonal arrangement with bright main lattice peaks surrounded by weaker CDW superstructure peaks, which arise from the CDW-coupled periodic lattice distortion \cite{Overhauser1971, Eichberger2010}.

\subsection*{Electron-laser spatial overlap}
We use single or double laser pulses to excite the sample under an angle of incidence of about 45 degrees. The laser spot size on the sample is 300\,$\upmu$m (FWHM), ensuring uniform illumination of the area probed by the electron beam. The spatial overlap between the electron and laser beams is found using the technique of ultrafast low-energy electron diffraction (ULEED) \cite{Vogelgesang2018}. Specifically, at low-fluence excitation, we optimize the electron-laser overlap by maximizing the laser-induced Debye-Waller effect, i.e., the transient suppression of diffraction spot intensities.

\subsection*{Generation of single- and double-pulse}
The optical quench of 1T-TaS$_{2}$ samples is conducted using either single or double laser pulses. Single pulses are isolated with a Pockels cell inside a laser amplifier (center wavelength 1030\,nm). Femtosecond excitation pulses of about 200\,fs duration (center wavelength 800\,nm, FWHM spectral width 7\,nm) are produced by an optical parametric amplifier. Double pulses with a pulse separation $|\tau| \le 2\,\textrm{ps}$ are generated by passing a single pulse through a Michelson interferometer. A half-inch polarizing beamsplitter cube is used to split the pulse, and a quarter-wave plate is used in each arm of the interferometer to optimize the total laser power output. The two output pulses are orthogonally polarized with a measured interferometric visibility of 3\,\%.

\subsection*{Reversal from the heterochiral to the monochiral state}
The duration of the annealing laser pulse train is controlled by a mechanical shutter that produces a 1\,s top-hat function with rising and falling edges spanning about 10\,ms. Given sufficient fluence, the annealing laser pulse train can completely reverse the 2D-heterochiral state to a monochiral state (cf. \Fig{fig:curve}c). This procedure is used for initializing each measurement. Incomplete reversal was observed if a sharper drop of the pulse train envelope was used.

\subsection*{Density functional theory simulations}
Density functional theory (DFT) simulations are performed for the CDW moir\'{e} superstructure using projector augmented wave (PAW) \cite{Blochl1994} potentials as implemented in the Vienna Ab-initio Simulation Package (VASP) \cite{Kresse1993, Kresse1996}. A 350 eV kinetic energy cutoff in the plane-wave expansion is used. The PAW pseudopotentials correspond to the valence-electron configuration 5d$^3$6s$^2$ for Ta and 3s$^2$3p$^4$ for S. The exchange-correlation is treated within the generalized gradient approximation, as parameterized by Perdew, Burke, and Ernzerhof \cite{Perdew1996}. The Coulomb correlations within the 5d shells of Ta are described using the spherically averaged DFT+$U$ method \cite{Dudarev1998}, where the Hamiltonian only depends on the difference between on-site interaction $U$ and Hund's exchange $J$, i.e., $U_{\textrm{eff}} = U-J$. We adopt the value of $U_{\textrm{eff}} = 2.0$\,eV for the Ta 5d orbitals \cite{Darancet2014}. Spin-orbit coupling is ignored in all simulations \cite{Ritschel2018}. For structural relaxation, the lattice constants $a = 43.68\,\AA$ (13 times of the $1\times1$ primitive cell lattice constant $3.36\,\AA$) and $c = 11.8\,\AA$ (twice the Ta-Ta interlayer distance $5.9\,\AA$) are fixed \cite{vonWitte2019, Spijkerman1997, Givens1977}, while the positions of 1014 atoms are relaxed by a $\Gamma$-centered $1\times1\times2$ k-point mesh until the maximum force on each atom becomes less than 10\,meV/$\AA$. A $\Gamma$-centered $3\times3\times2$ k-point mesh is used for calculating the self-consistent ground state charge and electronic band structures. The total energies are converged to 10$^{\textrm{-}4}$\,eV. In a separate calculation, The 1T-TaS$_2$ C-phase is simulated by a 156-atom AL-stacking unit cell using a $\Gamma$-centered $6\times6\times6$ k-point mesh \cite{Ritschel2018, Lee2019}.

\section*{Acknowledgments}
This work was funded by the European Research Council (ERC Starting Grant `ULEED', ID: 639119, ERC Advanced Grant `ULEEM', ID: 101055435), the Deutsche Forschungsgemeinschaft (SFB-1073, project A05) and ETH Z\"{u}rich (ETH Postdoctoral Fellowship, J.G.H.). This work utilized high-performance computing clusters from the Holland Computing Center of the University of Nebraska, which receives support from the UNL Office of Research and Economic Development, and the Nebraska Research Initiative. H.-T.J. acknowledges support from NSTC, NCHC, CINC-NTU, AS-iMATE-111-12, and CQT-NTHU-MOE, Taiwan. W.C.H. gratefully acknowledges funding supports from NSTC 112-2811-M-001-007 and 113-2811-M-007-008. The authors thank Gero Storeck, Benjamin Schr\"{o}der, Johannes Otto, Thomas Weitz, Matthias Kr\"{u}ger, and J\"{o}rg Malindretos for technical support and insightful discussions.


\clearpage
\onecolumngrid 

\input{Supp.tex}

\end{document}

%% file: Supp.tex

\setcounter{figure}{0}
\renewcommand{\thefigure}{S\arabic{figure}}

\providecommand{\FigsThree}[3]{Figs.~\ref{#1},~\ref{#2},~and~\ref{#3}}

\setcounter{table}{0}
\renewcommand{\thetable}{S\arabic{table}}

\setcounter{equation}{0}
\renewcommand{\theequation}{S\arabic{equation}}

\section*{Supplementary Information}

This supplement provides detailed information about the experimental observation and theoretical modeling of the heterochiral charge-density wave state (H-state). These include spot profile analysis, dynamical LEED simulations, as well as electronic structure, atomic lattice, and charge texture analysis based on density-functional theory (DFT) simulations. Additional data sets of diffraction intensity as a function of diffraction spot position and the probing electron energy (LEED I-V curves) are also provided.

\subsection{I. Spot profile analysis}

\begin{figure}[H] 
\centering
\includegraphics[width=\textwidth]{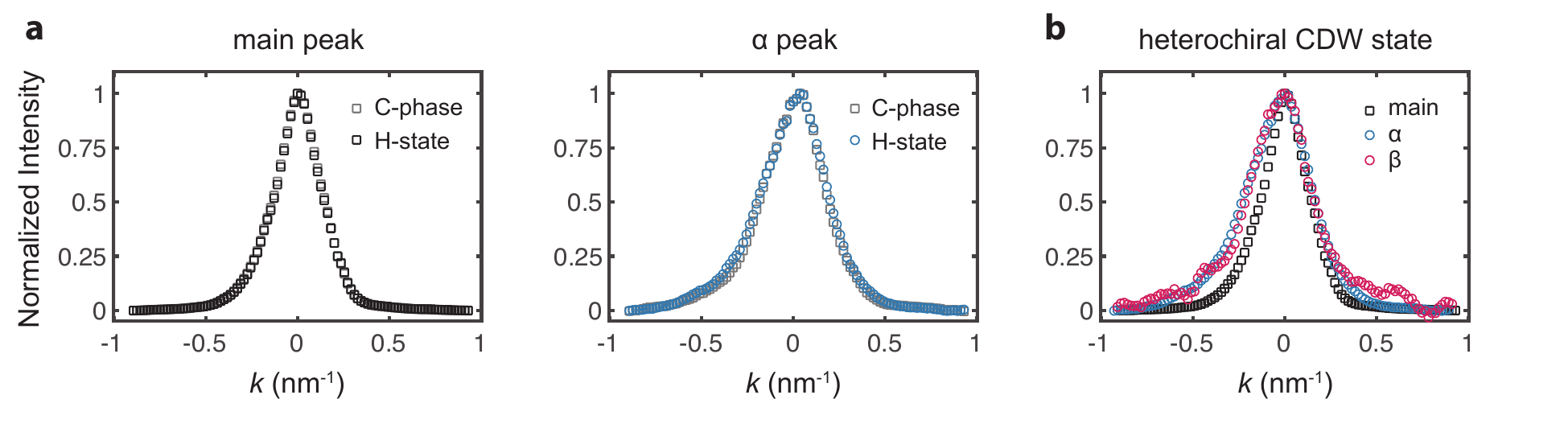}
\caption{Lineouts of the main lattice, $\alpha$, and $\beta$ peaks. 
\textbf{a}, Lineouts of the main lattice and $\alpha$ peaks, compared between the C-phase and the H-state. As the sample switches from the C-phase to the H-state after the optical quench, the diffraction intensities of the main lattice and $\alpha$ peak are slightly reduced, but there is negligible change in the linewidth. 
\textbf{b}, Lineout comparison in the H-state. The lineout of the emergent $\beta$ peak is similar to that of the $\alpha$ peak. Lineout angles for the $\alpha$ and $\beta$ peaks are along the radial directions, i.e., the main-$\alpha$ and main-$\beta$ directions (cf. main text Fig.\,1e, left column). The lineout angle for the main lattice peak is along the horizontal direction.
}
\label{fig:lineshape}
\end{figure}

Figure~\ref{fig:lineshape} presents the lineouts of the main lattice, $\alpha$, and $\beta$ peaks (cf. main text Fig.\,1e, right column). The lineout of the main lattice peak is along the horizontal direction, while the lineouts of the superstructure peaks are along the main-$\alpha$ or main-$\beta$ directions. In \Fig{fig:lineshape}a, we find negligible change in the linewidths of the main lattice and $\alpha$ peaks after the optical quench. In \Fig{fig:lineshape}b, lineouts of the main lattice, $\alpha$, and $\beta$ peaks in the H-state are compared, and their respective FWHMs are measured $\Delta k_{\text{main}}  = 0.28\,\textrm{nm}^{\textrm{-}1}$, $\Delta k_{\alpha}  = 0.40\,\textrm{nm}^{\textrm{-}1}$, and $\Delta k_{\beta}  = 0.38\,\textrm{nm}^{\textrm{-}1}$. The similar linewidths and lineouts of $\alpha$ and $\beta$ peaks evidence that the initial $\alpha$ chirality and the emergent $\beta$ chirality are both long-range order in the H-state. To estimate the lower bounds of $\alpha$ and $\beta$ domain sizes in the H-state, we approximate the diffraction lineouts in \Fig{fig:lineshape}b as Gaussians and calculate the corresponding correlation lengths using the deconvolution formula $\xi_{\alpha,\beta}=2\pi/\sqrt{\Delta k_{\alpha,\beta}^2-\Delta k_{\text{main}}^2}$, where we assume that the instrument response function follows the lineout of the main lattice peak. The retrieved minimum correlation lengths are $\xi_{\alpha}=22$\,nm and $\xi_{\beta}=24$\,nm for $\alpha$ and $\beta$ domains in the H-state.

\clearpage

\subsection{II. Experimental LEED I-V curves and dynamical LEED simulations} 

In this section, we investigate how the H-state signal varies with the diffraction spot position and the probing electron energy (LEED I-V curves). Dynamical LEED simulations are performed for comparison. 

\begin{figure}[!htbp]
\centering
\includegraphics[width=\textwidth]{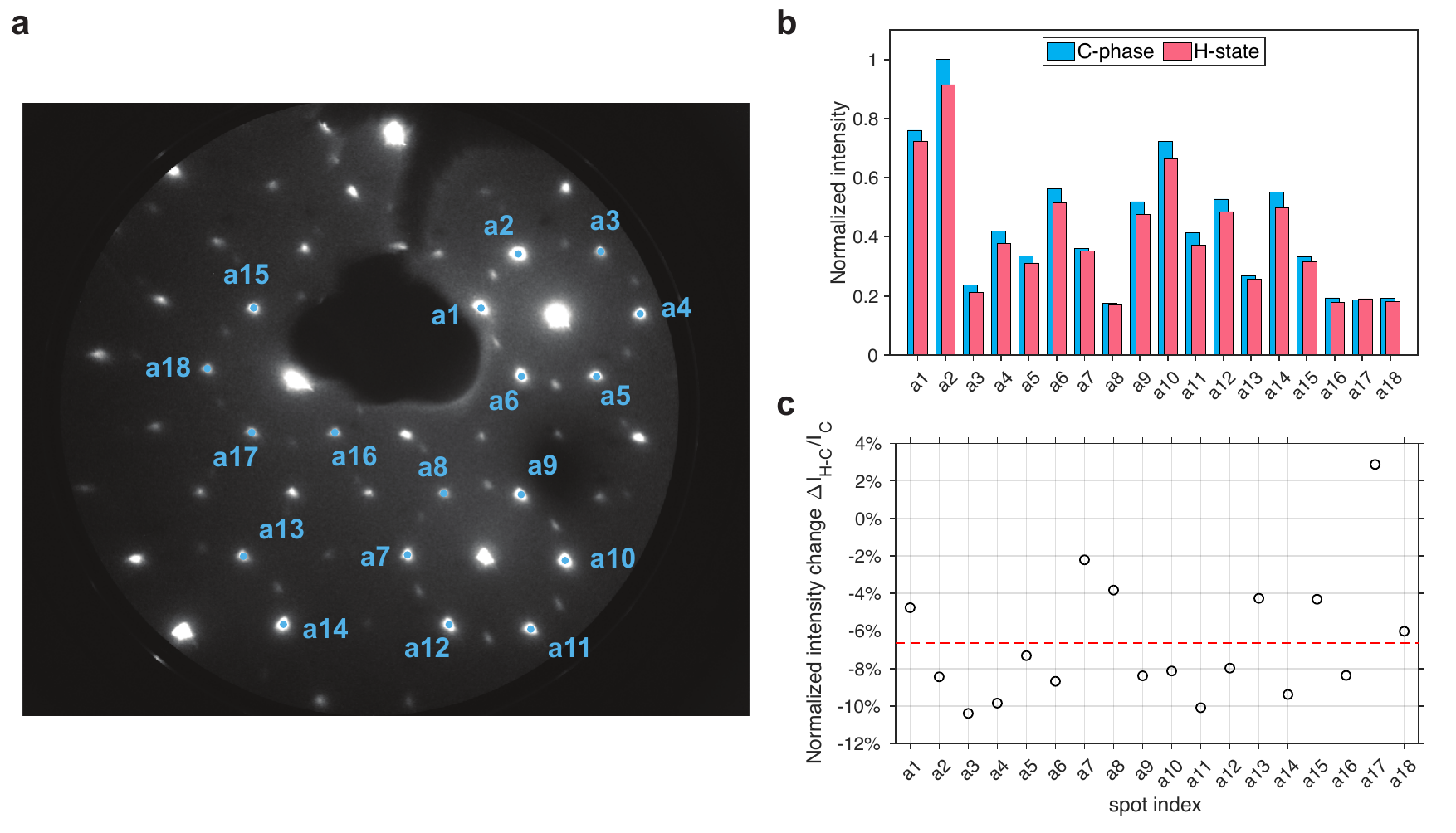}
\caption{Intensity of the $\alpha$ peak as a function of diffraction spot position.
\textbf{a}, Experimental LEED image of the H-state (cf. main text Fig.\,1d). Selected $\alpha$ peaks are labeled by spot indices a1 to a18.
\textbf{b}, Normalized $\alpha$-peak intensity as a function of diffraction spot position, compared between the C-phase and the H-state. All spot intensities decrease after the sample switches to the H-state.
\textbf{c}, Normalized intensity change of the $\alpha$ peak as a function of diffraction spot position. The average intensity change is -6.6\,\% (red dashed line).}
\label{fig:spot-alpha}
\end{figure}

\begin{figure}[!htbp]
\centering
\includegraphics[width=\textwidth]{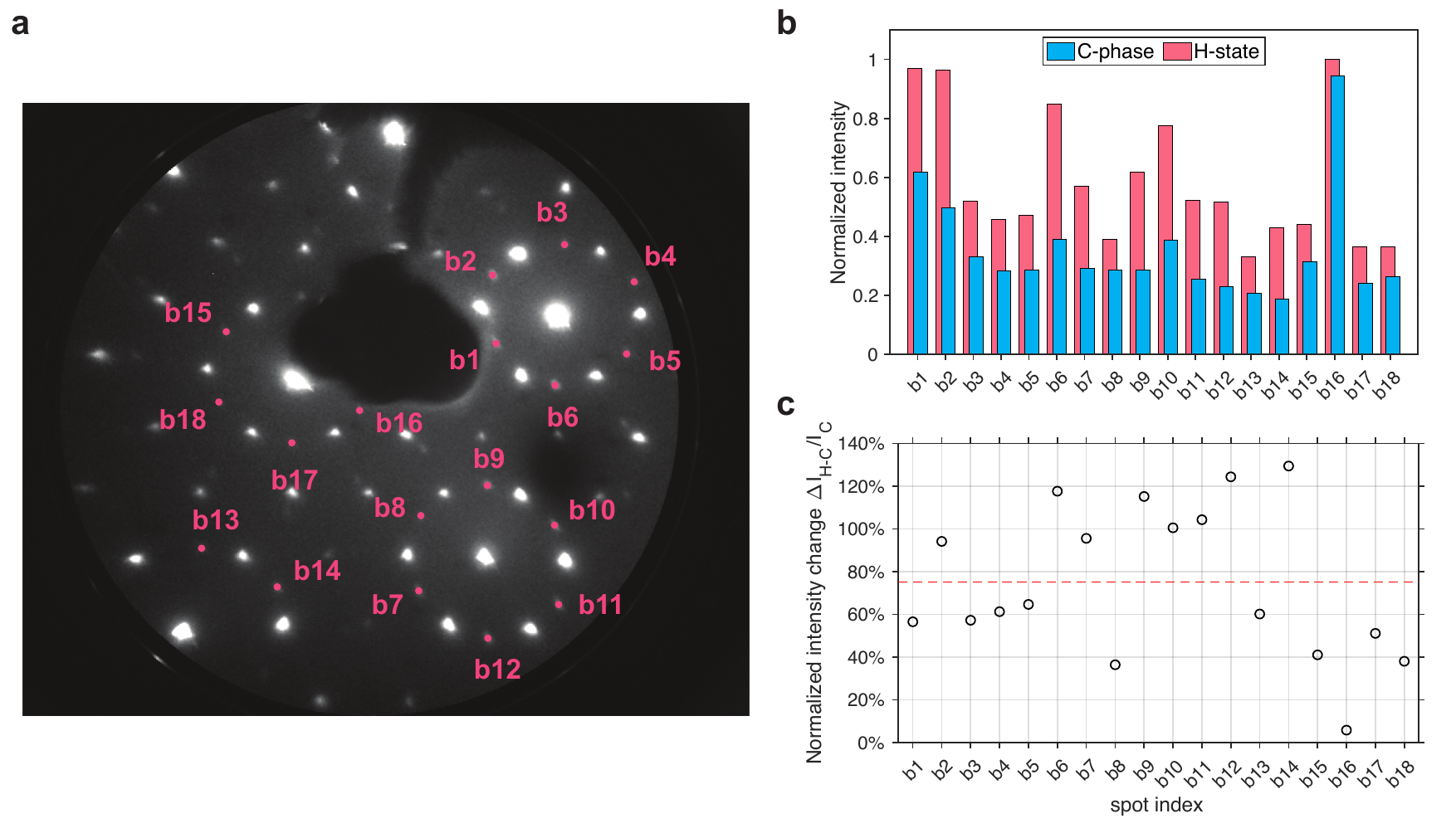}
\caption{Intensity of the $\beta$ peak as a function of diffraction spot position.
\textbf{a}, Experimental LEED image of the H-state (cf. main text Fig.\,1d). Selected $\beta$ peaks are labeled by spot indices b1 to b18.
\textbf{b}, Normalized $\beta$-peak intensity as a function of diffraction spot position, compared between the C-phase and the H-state. All spot intensities increase after the sample switches to the H-state.
\textbf{c}, Normalized intensity change of the $\beta$ peak as a function of diffraction spot position. The average intensity change (comparing to the background) is 75\,\% (red dashed line).
}
\label{fig:spot-beta}
\end{figure}

\begin{figure}[!htbp]
\centering
\caption{Experimental LEED I-V curves compared between the C-phase and the H-state.
\textbf{a}, Experimental LEED image of the H-state. The H-state is generated with a single-pulse quench at a fluence of 1.6\,mJ\,cm$^{-2}$. The probing electron energy is 100\,eV. Selected $\alpha$ and $\beta$ peaks are labeled by spot indices a1 to a14 and b1 to b14, respectively.
\textbf{b}, Experimental LEED I-V curves for $\alpha$ and $\beta$ peaks in the C-phase and the H-state.
\textbf{c}, Normalized intensity change of the $\alpha$ peak as a function of the probing electron energy and diffraction spot position.
\textbf{d}, Normalized intensity change of the $\beta$ peak as a function of the probing electron energy and diffraction spot position.
}
\includegraphics[width=\textwidth]{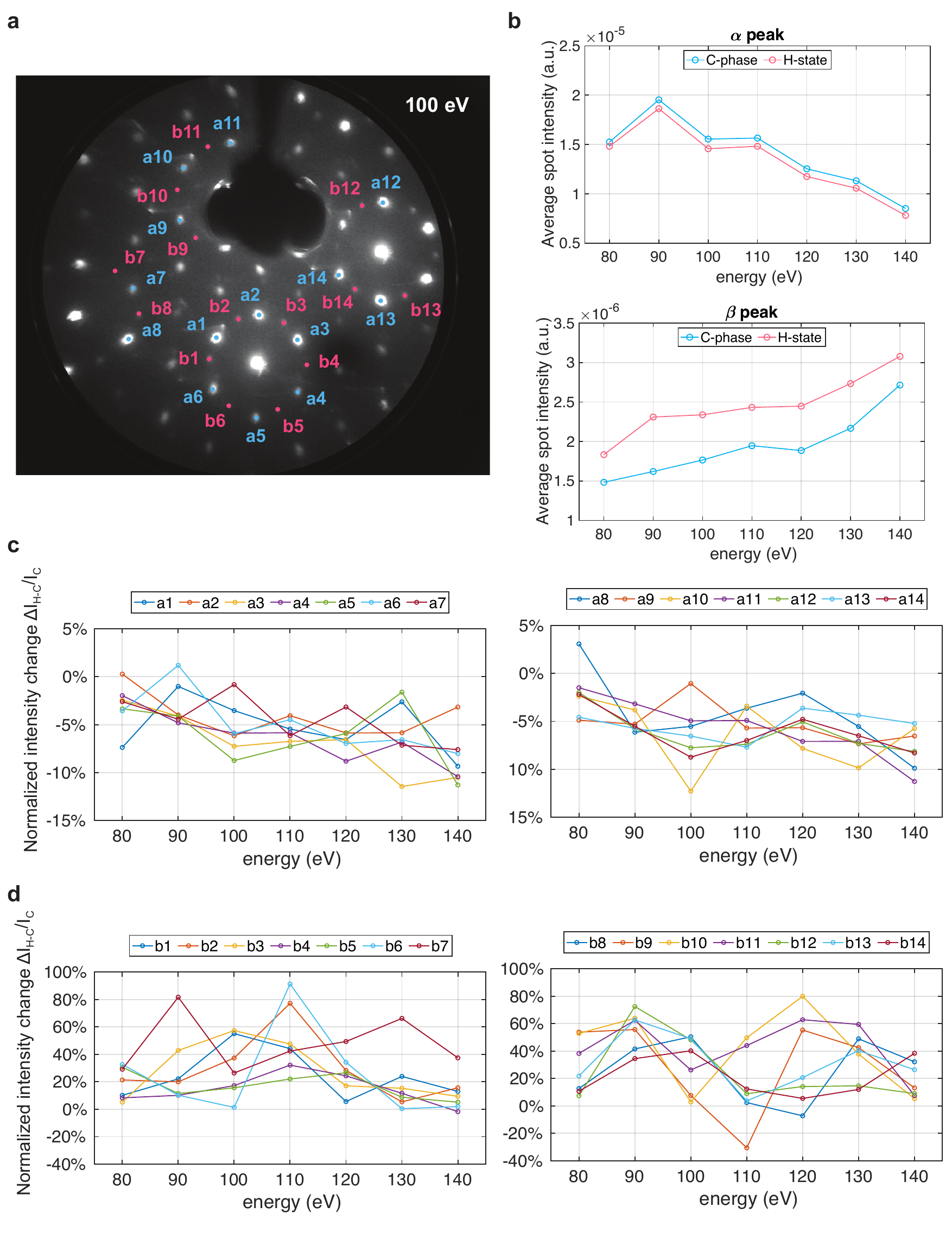}
\label{fig:LEED-IV}
\end{figure}

The H-state signal is quantified by the normalized intensity change, $\Delta I_{\text{H-C}}/I_{\text{C}} \equiv (I_{\text{H}}-I_{\text{C}})/I_{\text{C}}$, where $I_{\text{H,C}}$ is the diffraction spot intensity in the H-state or the C-phase, normalized to the total count of the respective LEED image. In \Fig{fig:spot-alpha}, the spot intensities of all $\alpha$ peaks decrease after the sample switches from the C-phase to the H-state. As a function of $\alpha$ spot position, $\Delta I_{\text{H-C}}/I_{\text{C}}$ fluctuates around -6.6\,\% with a standard deviation of 3.4\,\%. As a function of $\beta$ spot position, $\Delta I_{\text{H-C}}/I_{\text{C}}$ fluctuates around 75\,\% with a standard deviation of 36\,\% (\Fig{fig:spot-beta}). In both cases, the relative fluctuation is about 50\,\%. We note that $I_{\text{C}}$ of the $\beta$ peak takes the local background value, as $\beta$ peaks are absent in the C-phase. In \Fig{fig:LEED-IV}, $I_{\text{H,C}}$ and $\Delta I_{\text{H-C}}/I_{\text{C}}$ show only a mild modulation as a function of the probing electron energy.

\begin{figure}[!htbp]
\centering
\includegraphics[width=\textwidth]{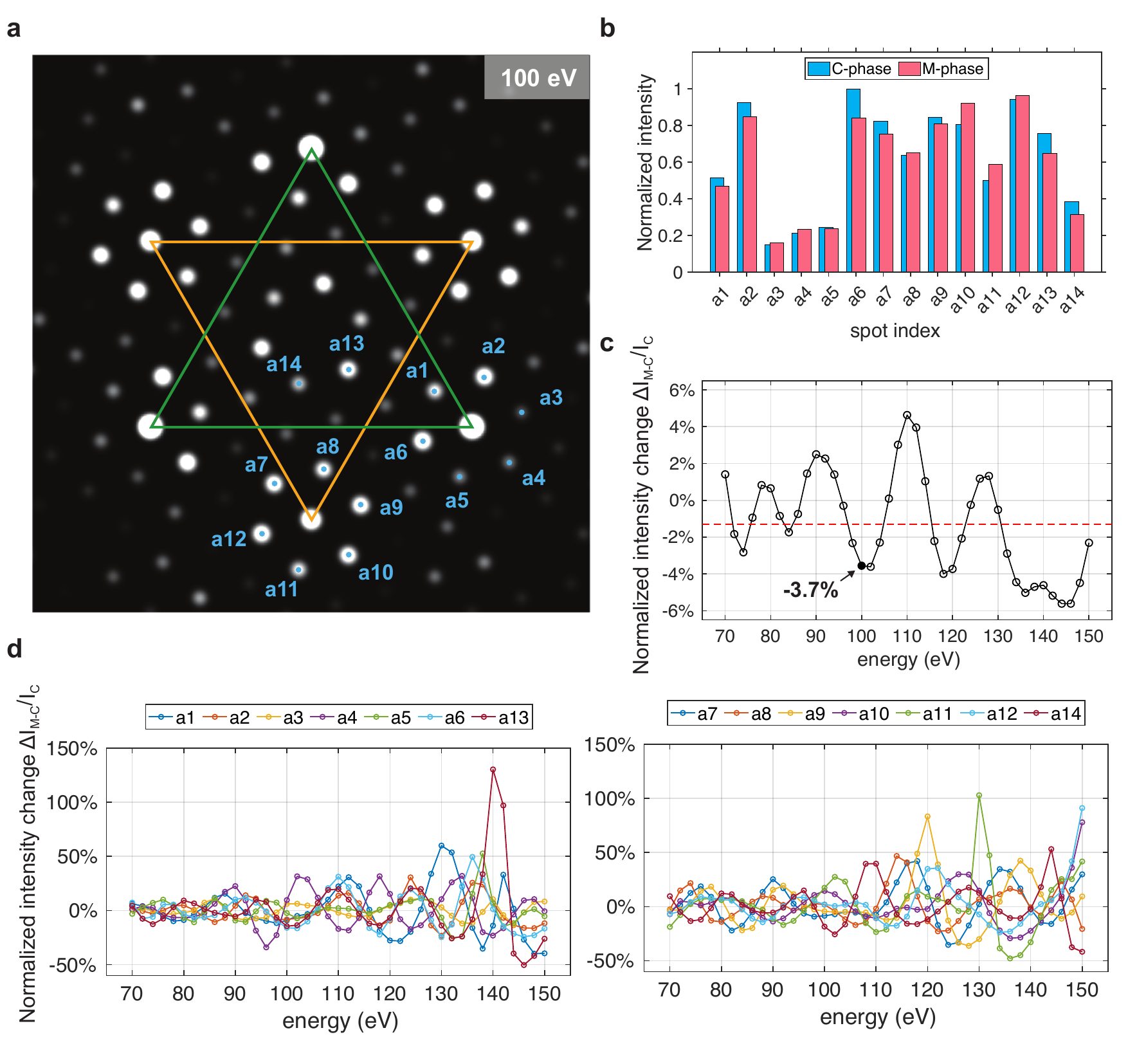}
\caption{Summary of the dynamical LEED simulations for two structures that differ in the subsurface layer.
\textbf{a}, Simulated LEED pattern of the C-phase structure. The LEED pattern is a graphical representation of the result of the dynamical LEED simulation, assuming 2D Gaussian spot profiles with the peak height determined by the simulation. Normal incidence is assumed for the probing electrons in the simulation. The effective three-fold symmetry divides the main lattice and CDW peaks into two groups (yellow and green triangles). Distinct first-order CDW peaks are labeled by spot indices a1 to a14.
\textbf{b}, Normalized intensity as a function of diffraction spot position, compared between the C- and M-phases. Due to multiple scattering effects, about half of the spots (a1, a2, a5, a6, a7, a9, a13, a14) have stronger intensity in the C-phase, while the other half (a3, a4, a8, a10, a11, a12) have stronger intensity in the M-phase.
\textbf{c}, Simulated LEED I-V curves of the subsurface layer contribution. The normalized intensity change, defined as $\Delta I_{\text{M-C}}/I_{\text{C}} \equiv (I_{\text{M}}-I_{\text{C}})/I_{\text{C}}$, quantifies the contribution of the subsurface layer in the total CDW diffraction intensity. The diffraction intensity in the M- and C-phases, $I_{\text{M,C}}$, is averaged over all 14 CDW spots (a1 to a14) at each probing electron energy. The intensity fluctuation as a function of the probing electron energy is a result of the multiple scattering effect. At 100\,eV, the subsurface layer contribution is 3.7\,\%.
\textbf{d}, Simulated LEED I-V curves of the subsurface layer contribution at individual diffraction spots. Here, $I_{\text{M,C}}$ is the intensity of individual diffraction spots. As a function of the probing electron energy, $\Delta I_{\text{M-C}}/I_{\text{C}}$ fluctuates between positive and negative values for all spots. This result is qualitatively different from our experimental observation in \Fig{fig:LEED-IV}c, where the intensity change takes only negative values.
}
\label{fig:LEED-sim}
\end{figure}

We investigate with dynamical LEED simulations how multiple scattering effects respond to a drastic structural change in the subsurface layer and compare the simulation results with experimental observations (\FigsThree{fig:spot-alpha}{fig:spot-beta}{fig:LEED-IV}). The simulation is based on our previous surface crystallographic analysis of the 1T-TaS$_{2}$ C-phase \cite{vonWitte2019}, using the TensErLEED package \cite{Blum2001}. We construct a two-layer C-phase structure and a hypothetical two-layer M-phase structure. The C-phase has the charge-density wave (CDW) induced periodic lattice distortion (PLD) on both surface and subsurface layers, while the M-phase has the surface layer with PLD and the subsurface layer without. By changing from the C-phase structure (surface: CDW, subsurface: CDW) to the M-phase structure (surface: CDW, subsurface: metallic), we simulate the effect of switching from the 1T-TaS$_{2}$ C-phase (surface: $\alpha$ chirality, subsurface: $\alpha$ chirality) to the H-state (surface: $\alpha$ chirality, subsurface: $\beta$ chirality), assuming the subsurface $\beta$-layer texture in Fig.\,3b of the main text. The simulation shows that a drastic structural change in the subsurface layer can result in large intensity fluctuations across all existing diffraction spots. Specifically, the spot intensity change $\Delta I_{\text{M-C}}/I_{\text{C}}$ takes both positive and negative values (\Fig{fig:LEED-sim}d). This result is qualitatively different from our experimental observation in \Fig{fig:LEED-IV}c, where all spot intensities of the $\alpha$ peak decrease after switching from the C-phase to the H-state.

\clearpage

\subsection{III. Mixed stacking order and the double-ring charge texture of the CDW \moire superstructure}

\begin{figure}[!htbp]
\centering
\includegraphics[width=\textwidth]{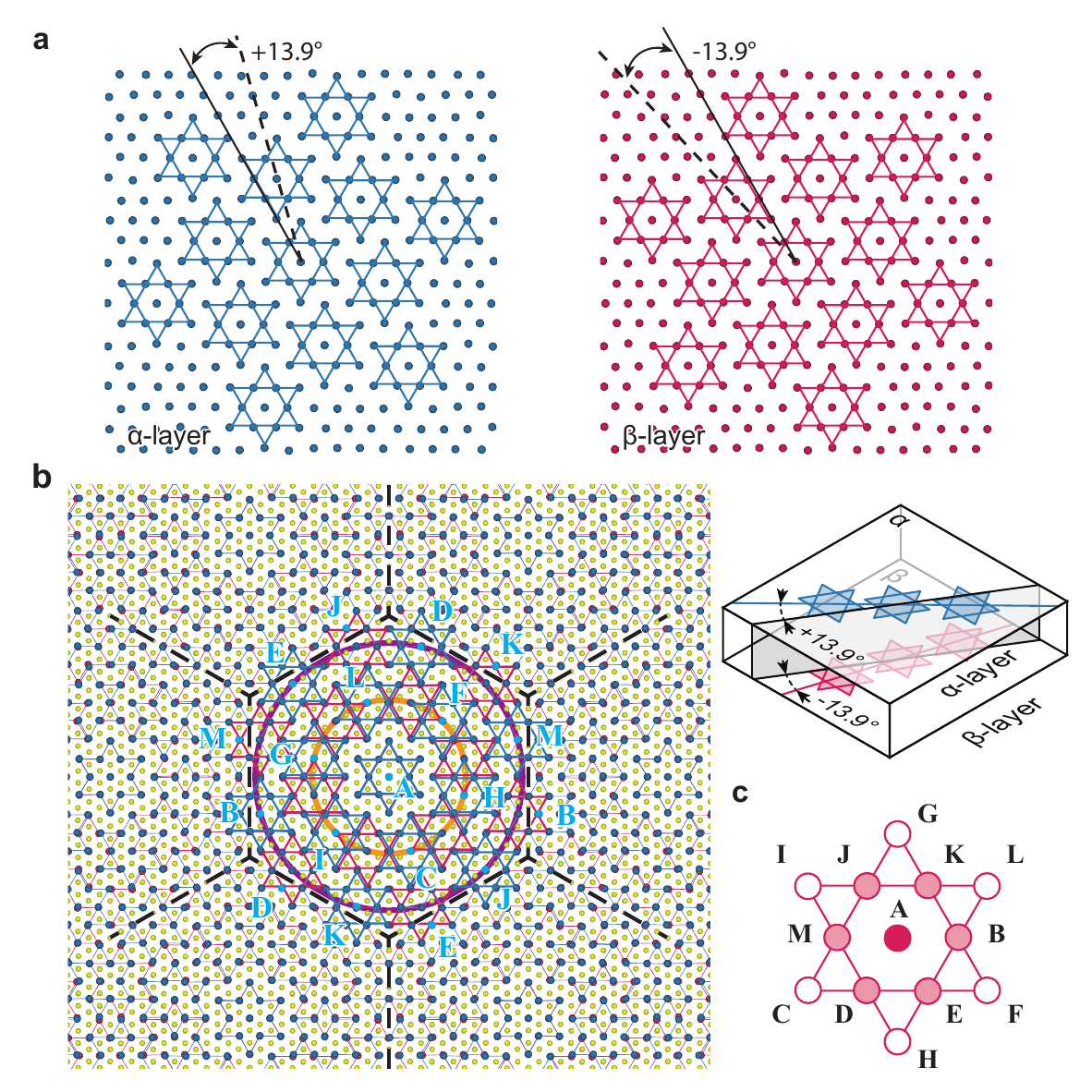}
\caption{Mixed stacking order of the CDW \moire superstructure.
\textbf{a}, Schematics of Ta super-hexagrams in the $\alpha$ and $\beta$ layers.
\textbf{b}, DFT simulated atomic structure of the light-induced CDW \moire superstructure in 1T-TaS$_{2}$. Blue and red dots represent Ta atoms in the $\alpha$ and $\beta$ layers, respectively. Yellow dots represent S atoms. The $\alpha$ layer (blue) is on the top and the $\beta$ layer (red) is at the bottom. Here the \moire unit cell (black dashed hexagon) is obtained through the construction of a Wigner-Seitz cell with the lattice constants $a = 43.68\,\AA$ and $c=11.8\,\AA$. For comparison, Ta hexagrams in the $\alpha$ and $\beta$ super-hexagram unit cells (cf. \textbf{a}) are highlighted by blue and red thick lines. The angle between the $\alpha$ and $\beta$ super-hexagrams give rise to a mixed stacking order. Namely, all 13 possible stacking types (denoted by capital letters A to M) are realized in the \moire unit cell. The A-stacking at the center has the highest Hartree energy. The I-, G-, L-, F-, H-, C-stackings at the inner ring (orange) of the super-hexagrams have the lowest Hartree energy. The K-, B-, E-, D-, M-, J-stackings at the outer ring (purple) of the super-hexagrams have the intermediate Hartree energy.
\textbf{c}, Illustration of the 13 stacking types according to Ref.~\cite{Lee2019}. The capital letters mark the center of the Ta hexagram on the top of each stacking pair (i.e., Ta hexagrams in the $\alpha$ layer). Darker filling color corresponds to higher electrostatic (Hartree) energy.
}
\label{fig:stacking}
\end{figure}

The CDW \moire superstructure has the $\alpha$ chirality in one layer and the $\beta$ chirality in the other (\Fig{fig:stacking}a). The $27.8^{\circ}$ rotational mismatch between CDWs in the $\alpha$ and $\beta$ layers gives rise to a commensurate \moire supercell of 1014 atoms (black dashed hexagon in \Fig{fig:stacking}b). In each layer, there are 13 Ta hexagrams, forming a super-hexagram in a fractal-type arrangement (\Fig{fig:stacking}a). The angle between $\alpha$ and $\beta$ super-hexagrams gives rise to a mixed stacking order, realizing all 13 possible stacking types (\Fig{fig:stacking}c) within the CDW \moire unit cell. Specifically, the A-stacking locates at the center of the super-hexagram, the six I-, G-, L-, F-, H-, C-stackings coincide with the inner ring of the super-hexagram, and the six K-, B-, E-, D-, M-, J-stackings reside at the outer ring (cf. \Fig{fig:stacking}b).

In our DFT simulations, the atomic positions are relaxed, while the lattice constants of the CDW \moire unit cell, $a = 43.68\,\AA$ (13 times of the $1\times1$ primitive cell lattice constant $3.36\,\AA$) and $c=11.8\,\AA$ (twice the Ta-Ta interlayer distance $5.90\,\AA$), are fixed. Prior to structural relaxation, the 13 Ta hexagrams in each of the $\alpha$ and $\beta$ layer of the CDW \moire unit cell are equivalent and the electron density should distribute evenly among them. However, in the presence of an electronic reconstruction, which occurs as a result of structural transformation (cf. \Fig{fig:relax}), the electron density redistributes into the inner and outer rings of the super-hexagrams, forming a double-ring charge texture.

To understand how the double-ring electron density distribution correlates with the atomic structure of the CDW \moire unit cell, we perform self-consistent energy calculations for the five nonequivalent stackings (A, B, K, H, C) of the standard 1T-TaS$_{2}$ structure. The five stackings are representative of all 13 stacking types due to the three-fold symmetry of 1T-TaS$_{2}$. The calculated electrostatic (Hartree) energy characterizes the Coulomb repulsion associated with each stacking type and may help to understand the stacking-dependent electron distributions in the CDW \moire superstructure.

The simulation results are summarized in \Table{tab:fixed}. The Hartree energy hierarchy $E_{\text{A}} > E_{\text{(B, D, J, K, E, M)}} > E_{\text{(H, I, L, C, G, F)}}$ is consistent with the double-ring charge texture. Namely, the highest Hartree energy of the local A-stacking order results in the depleted charge density at the center of the \moire unit cell. The lowest Hartree energy of the local H- and C-stacking order gives rise to the enhanced charge density at the inner ring of the super-hexagram, corresponding to the prominent rings shown in Fig.\,4c of the main text. The B- and K-stackings at the outer ring of the super-hexagram have the intermediate Hartree energy, which slightly suppress the charge density, forming a blurred ring of electron cloud.

\begin{table*}[!b]
\begin{ruledtabular}
\begin{tabular}{cccc}
Stacking type & Hartree energy, E (eV per hexagram) & Location in the super-hexagrams \\
\hline
A & 0 & center \\
\hline
B, D, J & -8.027 & outer ring \\
\hline
K, E, M & -8.155 & outer ring \\
\hline
H, I, L & -15.317 & inner ring \\
\hline
C, G, F & -15.422 & inner ring \\
\end{tabular}
\end{ruledtabular}
\caption{Electrostatic (Hartree) energy of standard 1T-TaS$_{2}$ lattice structures. Equivalent stacking types are grouped in the 1\textsuperscript{st} column. The Hartree energy of the A-stacking is 1055.407\,eV per hexagram. Energies of other stackings are referenced to the A-stacking. The DFT simulations are performed over a $\Gamma$-centered $6\times6\times4$ k-point mesh. In the simulation, the CDW unit cell lattice constants ($a=12.1147\,\AA$, $c=11.8\,\AA$) are fixed, while the atomic positions are relaxed.}
\label{tab:fixed}
\end{table*}

\clearpage

\subsection{IV. Emergent kagome structural order in the CDW \moire superstructure}

The DFT simulation for the CDW \moire superstructure starts with a two-layer supercell, each layer mimicking the surface-layer CDW superstructure in the 1T-TaS$_2$ C-phase \cite{vonWitte2019}. The CDW superstructure has two chiral orientations and a rotation angle of $13.9^{\circ}$ with respect to the main lattice (cf. \Fig{fig:stacking}a). We denote $13.9^{\circ}$ (clockwise) as the $\alpha$ chirality and -$13.9^{\circ}$ (counter-clockwise) as the $\beta$ chirality. The unit cell of the CDW \moire superstructure is built with the $\alpha$ chirality in the top layer (layer 1) and the $\beta$ chirality in the bottom layer (layer 2).

\begin{figure}[!htbp]
\centering
\includegraphics[width=\textwidth]{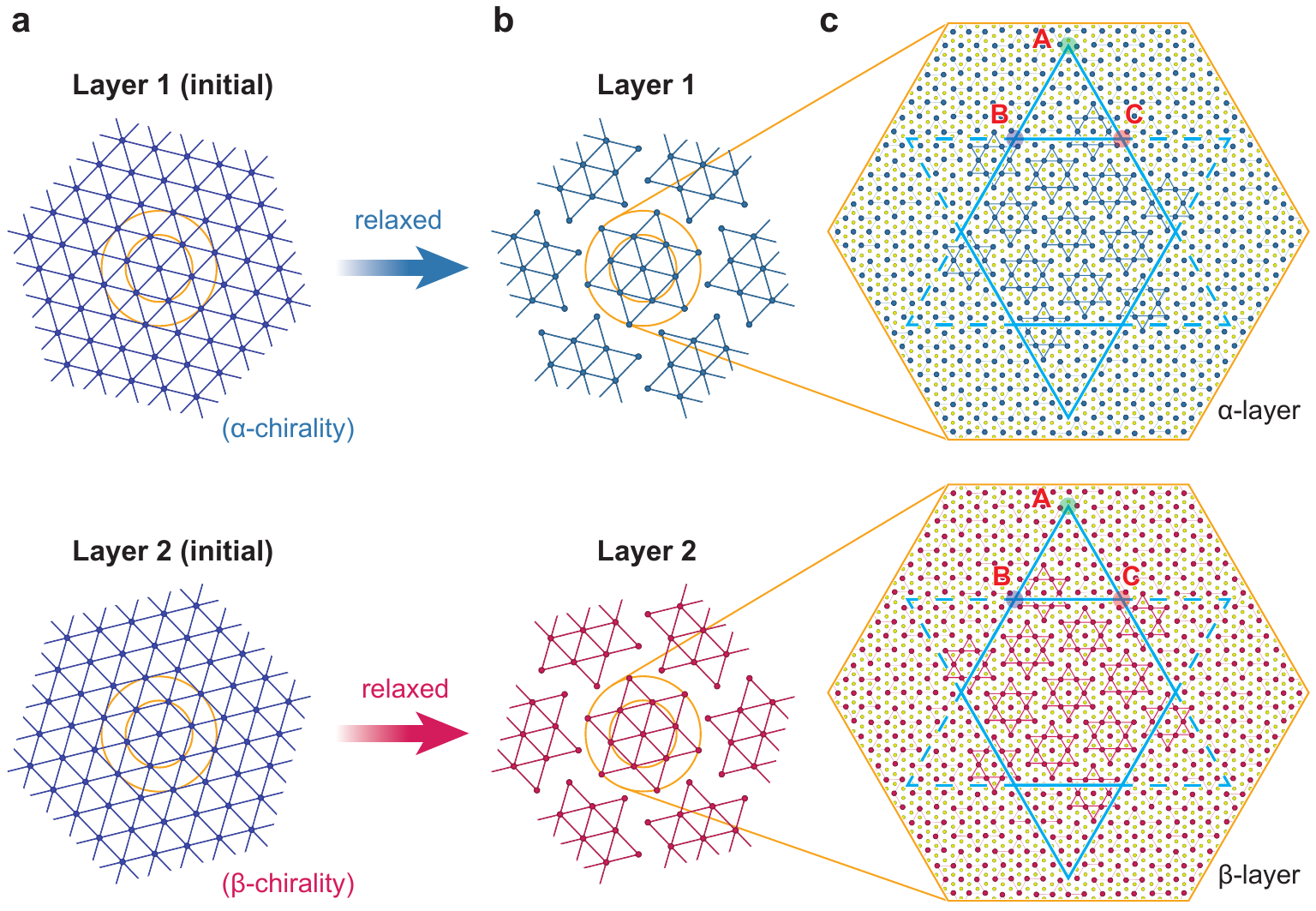}
\caption{Emergent kagome structural order in the CDW \moire superstructure.
\textbf{a}, The $(\sqrt{13}\times\sqrt{13}$)\textit{R}13.9$^{\circ}$ triangular lattice of the C-phase Ta superstructures. Only the Ta hexagram centers are shown (blue dots). Bonds of $12.1147\,\AA$ are plotted to illustrate the triangular lattice.
\textbf{b}, The $13\times13$ kagome lattice of the CDW \moire superstructure. The structural deformation caused by the coupling between $\alpha$ and $\beta$ layers clusters the 13 Ta hexagram centers into a super-hexagram. The super-hexagram has a three-fold symmetry, and bonds between neighboring super-hexagrams ($12.1182\,\AA$-$12.1242\,\AA$) are longer than those within the super-hexagram ($12.0999\,\AA$-$12.1171\,\AA$).
\textbf{c}, Closeup of the kagome superlattice with all atoms displayed. The kagome superlattice features corner-sharing triangles with three nonequivalent sites A, B, and C. Atoms near these three sites have orientations differing by $120^{\circ}$.
}
\label{fig:relax}
\end{figure}

\begin{figure}[!htbp]
\centering
\includegraphics[width=\textwidth]{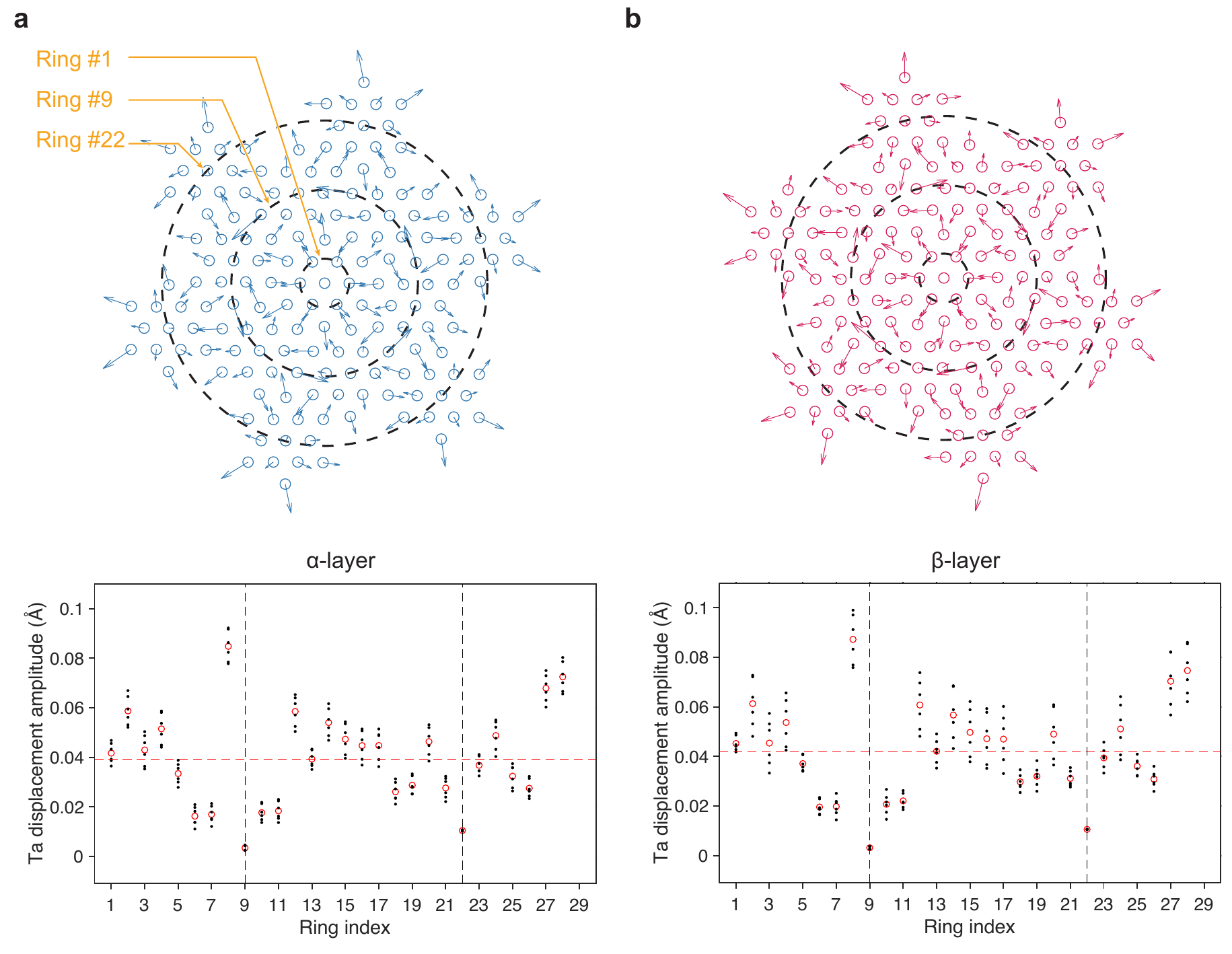}
\caption{Comparison between Ta positions in the CDW \moire superstructure and the C-phase.
\textbf{a}, Ta displacement from the C-phase structure in the $\alpha$ layer. (Top) The Ta displacements (depicted by arrows) can be grouped by 28 concentric rings according to the distance between the displaced Ta atom and the center of the $\alpha$ super-hexagram. As there are 169 Ta atoms within the super-hexagram unit cell, each ring is associated with 6 Ta atoms. The Ta atom at the center of the super-hexagram experiences no displacement. Ring \#9 and \#22 coincide with the inner and outer rings of the super-hexagram, where the Ta hexagram centers reside. (Bottom) Displacement amplitudes associated with each ring are shown by black dots, with the average value shown by red circles. The overall average of Ta displacement amplitude is around 0.04\,\AA\, (red dashed line), which is about 20\,\% of the CDW distortion in the C-phase.
\textbf{b}, Ta displacement from the C-phase structure in the $\beta$ layer. Displacements of Ta atoms in the $\alpha$ and $\beta$ layers have similar amplitudes but opposite sense of rotation with respect to the super-hexagram centers.
}
\label{fig:kagome}
\end{figure}

In the DFT simulation, the lattice constants of the \moire unit cell ($a=43.68\,\AA$, $c=11.8\,\AA$) are fixed, while the atomic positions are relaxed. The relaxed atomic structure experiences a structural transformation from a $(\sqrt{13}\times\sqrt{13}$)\textit{R}13.9$^{\circ}$ triangular superlattice to a $13\times13$ kagome superlattice (\Fig{fig:relax}b). The kagome superlattice features corner-sharing triangles with three nonequivalent sites (\Fig{fig:relax}c). Comparison between the relaxed atomic structure and the C-phase structure (AL-stacking) gives insight to the emergence of the kagome structural order. Namely, displacements of the Ta hexagram centers (\Fig{fig:kagome}) break the translational symmetry of Ta hexagram in both $\alpha$ and $\beta$ layers, transforming the Ta superstructure in each layer from $(\sqrt{13}\times\sqrt{13}$)\textit{R}13.9$^{\circ}$ to $13\times13$.

\clearpage

\subsection{V. Electronic band structures of the CDW \moire superstructure}

\begin{figure}[!htbp]
\centering
\includegraphics[width=\textwidth]{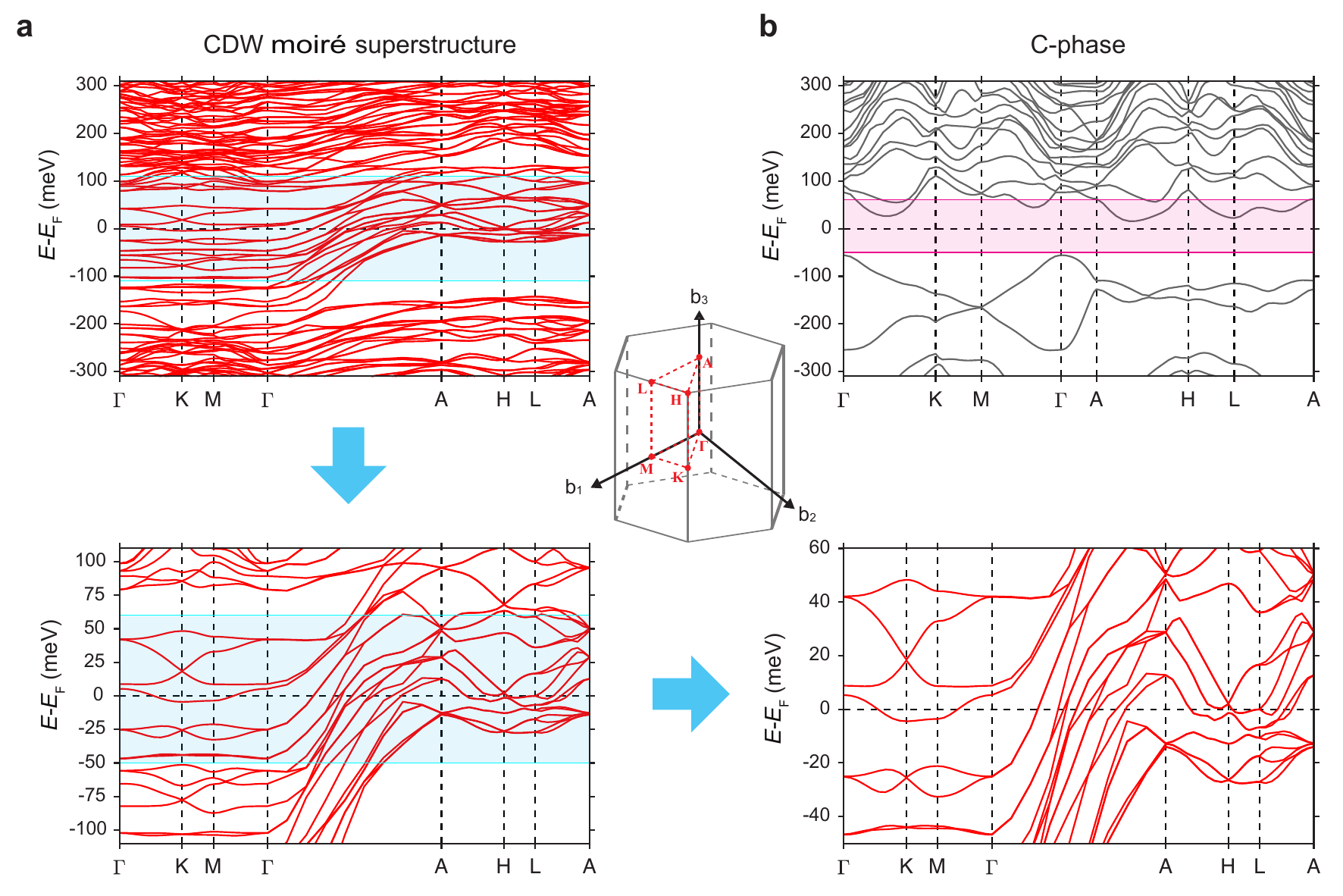}
\caption{Electronic band structures of the CDW \moire superstructure and the 1T-TaS$_{2}$ C-phase. \textbf{a}, Metallic band structure of the CDW \moire superstructure at different energy windows (blue shading). \textbf{b}, Insulating band structure of the 1T-TaS$_{2}$ C-phase (i.e., AL-stacking). An energy gap is present from -50\,meV to 60\,meV around $\Ef$ (red shading). (Middle) The 1\textsuperscript{st} Brillouin zone of the CDW \moire superstructure. Red dots represent high-symmetry k-points. The aspect ratio of the Brillouin zone ($\approx 0.3$) is adjusted for better visualization of the sampling paths ($\Gamma$-K-M-$\Gamma$-A-H-L-A).}
\label{fig:band}
\end{figure}

Here we present the electronic band structures obtained from DFT simulations, using the Vienna Ab initio Simulation Package \cite{Kresse1996, Kresse1999, Blochl1994} with an on-site $U_{\textrm{eff}}$ of 2.0 eV for Ta 5d orbitals \cite{Darancet2014} in the generalized gradient approximation of the Perdew Burke Ernzerhof (PBE) form \cite{Perdew1996}. After relaxing the atomic positions (cf. \Fig{fig:relax}), we perform self-consistent electronic structure calculations with a $3\times3\times2$ $\Gamma$-centered Monkhorst-Pack k-point mesh. The cutoff energy for the plane-wave basis is 400 eV. The band structure shown in \Fig{fig:band}a is calculated along paths connecting the 6 high-symmetry k-points in the 1\textsuperscript{st} Brillouin zone of the CDW \moire unit cell. In stark contrast to the insulating band structure of the C-phase (\Fig{fig:band}b), the CDW \moire superstructure exhibits multiple flat bands and Dirac cones within $\pm100$\,meV of the Fermi energy ($\Ef$) due to the mixed stacking order (cf. \Fig{fig:stacking}), the structural transformation (cf. \Fig{fig:relax}), and the electronic reconstruction. Emergent metallicity is evidenced by the density-of-states peak at $\Ef$ (cf. main text Fig.\,4d) as well as the emergent bands. On the $\Gamma$-K-M plane, at 10\,meV above $\Ef$ there is a kagome electronic subsystem that can facilitate in-plane conductivity. On the A-H-L plane, there are several bands crossing $\Ef$, allowing for non-vertical interlayer hopping. Along the $\Gamma$-A direction, the dispersive bands reflect an out-of-plane metallic character.

\section*{Supplementary References}

%% file: Combined.bbl
\begin{thebibliography}{99}



	
	



\bibitem{Bao2022} C. Bao, P. Tang, D. Sun, and S. Zhou, \textit{Light-induced emergent phenomena in 2D materials and topological materials}, Nat. Rev. Phys. \textbf{4}, 33 (2022). 

\bibitem{Torre2021} A. de la Torre, D. M. Kennes, M. Claassen, S. Gerber, J. W. McIver, and M. A. Sentef, \textit{Colloquium: Nonthermal pathways to ultrafast control in quantum materials}, Rev. Mod. Phys. \textbf{93}, 041002 (2021).  

\bibitem{Basov2017} D. N. Basov, R. D. Averitt, and D. Hsieh, \textit{Towards properties on demand in quantum materials}, Nat. Mater. \textbf{16}, 1077 (2017). 
   	
\bibitem{Stojchevska2014} L. Stojchevska, I. Vaskivskyi, T. Mertelj, P. Kusar, D. Svetin, S. Brazovskii, and D. Mihailovic, \textit{Ultrafast switching to a stable hidden quantum state in an electronic crystal}, Science \textbf{344}, 177 (2014). 

\bibitem{Horstmann2020} J. G. Horstmann, H. B\"{o}ckmann, B. Wit., F. Kurtz, G. Storeck, and C. Ropers, \textit{Coherent control of a surface structural phase transition}, Nature \textbf{583}, 232 (2020). 

\bibitem{Rini2007} M. Rini, R. Tobey, N. Dean, J. Itatani, Y. Tomioka, Y. Tokura, R. W. Schoenlein, and A. Cavalleri, \textit{Control of the electronic phase of a manganite by mode-selective vibrational excitation}, Nature \textbf{449}, 72 (2007).
    
\bibitem{Vaskivskyi2016} I. Vaskivskyi, I.A. Mihailovic, S. Brazovskii, J. Gospodaric, T. Mertelj, D. Svetin, P. Sutar, and D. Mihailovic, \textit{Fast electronic resistance switching involving hidden charge density wave states}, Nat. Commun. \textbf{7}, 11442 (2016).	

\bibitem{Burri2025}
C. Burri, N. Hua, D. F. Sanchez, W. Hu, H. G. Bell, R. Venturini, S.-W. Huang,
A. G. McConnell, F. Dizdarevic, A. Mraz, D. Svetin, B. Lipov\v{s}ek, M. Topi\v{c},
D. Kazazis, G. Aeppli, D. Grolimund, Y. Ekinci, D. Mihailovi\'{c}, and S. Gerber,
\textit{Imaging of electrically controlled van der Waals layer stacking in 1T-TaS$_2$},
Nat. Commun. \textbf{16}, 10296 (2025).



\bibitem{Gao2022} F. Y. Gao, Z. Zhang, Z. Sun, L. Ye, Y. -H. Cheng, Z. -J. Liu, J. G. Checkelsky, E. Baldini, and K. A. Nelson, \textit{Snapshots of a light-induced metastable hidden phase driven by the collapse of charge order}, Sci. Adv. \textbf{8}, eabp9076 (2022).

\bibitem{Svetin2017} D. Svetin, I. Vaskivskyi, S. Brazovskii, and D. Mihailovic, \textit{Three-dimensional resistivity and switching between correlated electronic states in 1T-TaS$_{2}$}, Sci. Rep. \textbf{7}, 46048 (2017).

\bibitem{vonWitte2019} G. von Witte, T. Ki\ss linger, J. G. Horstmann, K. Rossnagel, M. A. Schneider, C. Ropers, and L. Hammer, \textit{Surface structure and stacking of the commensurate ($\sqrt{13}\times\sqrt{13}$)R13.9$^{\circ}$ charge density wave phase of 1T-TaS$_{2}$(0001)}, Phys. Rev. B \textbf{100}, 155407 (2019). 
	
\bibitem{Ma2016} L. Ma, C. Ye, Y. Yu, X. F. Lu, X. Niu, S. Kim, D. Feng, D. Tom\'{a}nek, Y. -W. Son, X. H. Chen, and Y. Zhang, \textit{A metallic mosaic phase and the origin of Mott-insulating state in 1T-TaS$_{2}$}, Nat. Commun. \textbf{7}, 10956 (2016). 

\bibitem{Cho2016} D. Cho, S. Cheon, K. -S. Kim, S. -H. Lee, Y. -H. Cho, S. -W. Cheong, and H. W. Yeom, \textit{Nanoscale manipulation of the Mott insulating state coupled to charge order in 1T-TaS$_2$}, Nat. Commun. \textbf{7}, 10453 (2016). 

\bibitem{Gerasimenko2019} Y. A. Gerasimenko, P. Karpov, I. Vaskivskyi, S. Brazovskii, and D. Mihailovic, \textit{Intertwined chiral charge orders and topological stabilization of the light-induced state of a prototypical transition metal dichalcogenide}, npj Quantum Mater. \textbf{4}, 32 (2019). 

\bibitem{Vaskivskyi2015} I. Vaskivskyi, J. Gospodaric, S. Brazovskii, D. Svetin, P. Sutar, E. Goreshnik, I. A. Mihailovic, T. Mertelj, and D. Mihailovic, \textit{Controlling the metal-to-insulator relaxation of the metastable hidden quantum state in 1T-TaS$_{2}$}, Sci. Adv. \textbf{1}, e1500168 (2015).
		
\bibitem{Stahl2020} Q. Stahl, M. Kusch, F. Heinsch, G. Garbarino, N. Kretzschmar, K. Hanff, K. Rossnagel, J. Geck, and T. Ritsche, \textit{Collapse of layer dimerization in the photo-induced hidden state of 1T-TaS$_{2}$}, Nat. Commun. \textbf{11}, 1247 (2020). 

\bibitem{Vogelgesang2018} S. Vogelgesang, G. Storeck, J. G. Horstmann, T. Diekmann, M. Sivis, S. Schramm, K. Rossnagel, S. Sch\"{a}fer, and C. Ropers, \textit{Phase ordering of charge density waves traced by ultrafast low-energy electron diffraction}, Nat. Phys. \textbf{14}, 185 (2018).  

\bibitem{Ravnik2018} J. Ravnik, I. Vaskivskyi, T. Mertelj, and D. Mihailovic, \textit{Real-time observation of the coherent transition to a metastable emergent state in 1T-TaS$_{2}$}, Phys. Rev. B \textbf{97}, 075304 (2018). 

\bibitem{Song2022} X. Song, L. Liu, Y. Chen, H. Yang, Z. Huang, B. Hou, Y. Hou, X. Han, H. Yang, Q. Zhang, T. Zhang, J. Zhou, Y. Huang, Y. Zhang, H. -J. Gao, and Y. Wang, \textit{Atomic-scale visualization of chiral charge density wave superlattices and their reversible switching}, Nat. Commun. \textbf{13}, 1843 (2022).

\bibitem{Liu2023} L. Liu, X. Song, J. Dai, H. Yang, Y. Chen, X. Huang, Z. Huang, H. Ji, Y. Zhang, X. Wu, J. -T. Sun, Q. Zhang, J. Zhou, Y. Huang, J. Qiao, W. Ji, H. -J. Gao, and Y. Wang, \textit{Unveiling electronic behaviors in heterochiral charge-density-wave twisted stacking materials with 1.25 nm unit dependence}, ACS Nano \textbf{17}, 2702 (2023).

\bibitem{Ohta2021} S. Ohta, S. Kobayashi, A. Nomura, and H. Sakata, \textit{Electronic state modulation of the Star of David lattice by stacking of $\sqrt{13}\times\sqrt{13}$ domains in 1T-TaSe$_2$}, Phys. Rev. B \textbf{104}, 155433 (2021).
	
\bibitem{Stojchevska2018} L. Stojchevska, P. \v{S}utar, E. Goreshnik, D. Milhailovic, and T. Mertelj, \textit{Stability of the light-induced hidden charge density wave state within the phase diagram of 1T-TaS$_{\textrm{2-x}}$Se$_{\textrm{x}}$}, Phys. Rev. B \textbf{98}, 195121 (2018).


\bibitem{Storeck2021} G. Storeck, K. Rossnagel, and C. Ropers, \textit{Ultrafast spot-profile LEED of a charge-density wave phase transition}, Appl. Phys. Lett. \textbf{118}, 221603 (2021).

\bibitem{Maklar2023} J. Maklar, J. Sarkar, S. Dong, Y. A. Gerasimenko, T. Pincelli, S. Beaulieu, P. S. Kirchmann, J. A. Sobota, S. Yang, D. Leuenberger, R. G. Moore, Z. -X. Shen, M. Wolf, D. Mihailovic, R. Ernstorfer, and L. Rettig , \textit{Coherent light control of a metastable hidden state}, Sci. Adv. 9, eadi4661 (2023).


	
\bibitem{Zong2018} A. Zong, X. Shen, A. Kogar, L. Ye, C. Marks, D. Chowdhury, T. Rohwer, B. Freelon, S. Weathersby, R. Li, J. Yang, J. Checkelsky, X. Wang, and N. Gedik, \textit{Ultrafast manipulation of mirror domain walls in a charge density wave}, Sci. Adv. \textbf{4}, eaau5501 (2018). 

\bibitem{Delatorre2025}
A. de la Torre, Q. Wang, Y. Masoumi, B. Campbell, J. V. Riffle, D. Balasundaram, P. M. Vora,
J. P. C. Ruff, G. A. Fiete, S. M. Hollen, and K. W. Plumb,
\textit{Dynamic phase transition in 1T-TaS$_{2}$ via a thermal quench},
Nat. Phys. \textbf{21}, 1267--1274 (2025).
    
\bibitem{Vogelgesang2018thesis} S. Vogelgesang, \textit{Ultrafast low-energy electron diffraction at surfaces}, PhD Thesis, University of G\"{o}ttingen (2018).

\bibitem{Ravnik2023} J. Ravnik, J. Vodeb, Y. Vaskivskyi, M. Diego, R. Venturini, Y. Gerasimenko, V. Kabanov, A. Kranjec, and D. Mihailovic, \textit{Chiral domain dynamics and transient interferences of mirrored superlattices in nonequilibrium electronic crystals}, Sci. Rep. \textbf{13}, 19622 (2023).




\bibitem{Lee2019} S. -H. Lee, J. S. Goh, and D. Cho, \textit{Origin of the insulating phase and first-order metal-insulator transition in 1T-TaS$_{2}$}, Phys. Rev. Lett. \textbf{122}, 106404 (2019).

\bibitem{Ritschel2018} T. Ritschel, H. Berger, and J. Geck, \textit{Stacking-driven gap formation in layered 1T-TaS$_{2}$}, Phys. Rev. B \textbf{98}, 195134 (2018).
	
\bibitem{Nicholson2022} C. W. Nicholson, F. Petocchi, B. Salzmann, C. Witteveen, M. Rumo, G. Kremer, F. O. von Rohr, P. Werner, and C. Monney, \textit{Modified interlayer stacking and insulator to correlated-metal transition driven by uniaxial strain in 1T-TaS$_{2}$}, arXiv:2204.05598 (2022).

\bibitem{Salzmann2023} B. Salzmann, E. Hujala, C. Witteveen, B. Hildebrand, H. Berger, F. O. von Rohr, C. W. Nicholson, and C. Monney, \textit{Observation of the metallic mosaic phase in 1T-TaS$_{2}$ at equilibrium}, Phys. Rev. Mater. \textbf{7}, 064005 (2023).

\bibitem{Butler2020} C. J. Butler, M. Yoshida, T. Hanaguri, and Y. Iwasa, \textit{Mottness versus unit-cell doubling as the driver of the insulating state in 1T-TaS$_{2}$}, Nat. Commun. \textbf{11}, 2477 (2020).

\bibitem{Lee2021} J. Lee, K. -H. Jin, and H. W. Yeom, \textit{Distinguishing a Mott Insulator from a trivial insulator with atomic adsorbates}, Phys. Rev. Lett. \textbf{126}, 196405 (2021).

\bibitem{Petocchi2022} F. Petocchi, C. W. Nicholson, B. Salzmann, D. Pasquier, O. V. Yazyev, C. Monney, and P. Werner, \textit{Mott versus hybridization gap in the low-temperature phase of 1T-TaS$_{2}$}, Phys. Rev. Lett. \textbf{129}, 016402 (2022).

\bibitem{Fei2022} Y. Fei, Z. Wu, W. Zhang, and Y. Yin, \textit{Understanding the Mott insulating state in 1T-TaS$_{2}$ and 1T-TaSe$_{2}$}, AAPPS Bull. \textbf{32}, 20 (2022).

\bibitem{Smith2001}
N. V. Smith, \textit{Classical generalization of the Drude formula for the optical conductivity},
Phys. Rev. B \textbf{64}, 155106 (2001).


\bibitem{Deutschlander2015} S. Deutschl\"{a}nder, P. Dillmann, G. Maret, and P. Keim, \textit{Kibble-Zurek mechanism in colloidal monolayers}, PNAS \textbf{112}, 6925 (2015).

\bibitem{Eggebrecht2017} T. Eggebrecht, M. M\"{o}ller, J. G. Gatzmaan. N. R. da Silva, A. Feist, U. Martens, H. Ulrichs, M. M\"{u}nzenberg, C. Ropers, and S. Sch\"{a}fer, \textit{Light-induced metastable magnetic texture uncovered by in situ Lorentz microscopy}, Phys. Rev. Lett. \textbf{118}, 097203 (2017).

\bibitem{Storeck2020} G. Storeck, J. G. Horstmann, T. Diekmann, S. Vogelgesang, G. von Witte, S. V. Yalunin, K. Rossnagel, and C. Ropers, \textit{Structural dynamics of incommensurate charge-density waves tracked by ultrafast low-energy electron diffraction}, Struct. Dyn. \textbf{7}, 034304 (2020).


\bibitem{Shallcross2010} S. Shallcross, S. Sharma, E. Kandelaki, and O. A. Pankratov, \textit{Electronic structure of turbostratic graphene}, Phys. Rev. B \textbf{81}, 1 (2010). 

\bibitem{Tilak2023} N. Tilak, M. Altvater, S. -H. Hung, C. -J. Won, G. Li, T. Kaleem, S. -W. Cheong, C. -H. Chung, H. -T. Jeng, and E. Y. Andrei, \textit{Revealing the charge density wave proximity effect in graphene on 1T-TaS$_{2}$}, arXiv:2311.10606  (2023).
	
\bibitem{Carr2020} S. Carr, S. Fang, and E. Kaxiras, \textit{Electronic-structure methods for twisted moir\'{e} layers}, Nat. Rev. Mater. \textbf{5}, 749 (2020).

\bibitem{Cao2018-1} Y. Cao, V. Fatemi, S. Fang, K. Watanabe, T. Taniguchi, E. Kaxiras, and P. Jarillo-Herrero, \textit{Unconventional superconductivity in magic-angle graphene superlattices}, Nature \textbf{556}, 43 (2018).

\bibitem{Cao2018-2} Y. Cao, V. Fatemi, A. Demir, S. Fang, S. L. Tomarken, J. Y. Luo, J. D. Sanchez-Yamagishi, K. Watanabe, T. Taniguchi, E. Kaxiras, R. C. Ashoori, P. Jarillo-Herrero, \textit{Correlated insulator behavior at half-filling in magic-angle graphene superlattices}, Nature \textbf{556}, 80 (2018).

\bibitem{Yin2022}  J. -X. Yin, B. Lian, and M. Z. Hasan, \textit{Topological kagome magnets and superconductors}, Nature \textbf{612}, 647 (2022).
	


\bibitem{Ishioka2010} J. Ishioka, Y. H. Liu, K. Shimatake, T. Kurosawa, K. Ichimura, Y. Toda, M. Oda, and S. Tanda, \textit{Chiral charge-density waves}, Phys. Rev. Lett. \textbf{105}, 176401 (2010).	

\bibitem{Overhauser1971} A. Overhauser, \textit{Observability of charge-density waves by neutron diffraction}, Phys. Rev. B \textbf{10}, 3173 (1971).
	
\bibitem{Eichberger2010} M. Eichberger, H. Sch\"{a}fer, M. Krumova, M. Beyer, J. Demsar, H. Berger, G. Moriena, G. Sciaini, and R. J. D. Miller, \textit{Snapshots of cooperative atomic motions in the optical suppression of charge density waves}, Nature \textbf{468}, 799 (2010).
		
\bibitem{Blochl1994} P. E. Bl\"{o}chl, \textit{Projector augmented-wave method}, Phys. Rev. B \textbf{50}, 17953 (1994).
	
\bibitem{Kresse1993} G. Kresse and J. Hafner, \textit{Ab initio molecular dynamics for liquid metals}, Phys. Rev. B \textbf{47}, 558 (1993).
	
\bibitem{Kresse1996} G. Kresse and J. Furthm\"{u}ller, \textit{Efficient iterative schemes for ab initio total-energy calculations using a plane-wave basis set}, Phys. Rev. B \textbf{54}, 11169 (1996).
	
\bibitem{Perdew1996} J. P. Perdew, K. Burke, and M. Ernzerhof, \textit{Generalized Gradient Approximation Made Simple}, Phys. Rev. Lett. \textbf{77}, 3865 (1996).
	
\bibitem{Dudarev1998} S. L. Dudarev, G. A. Botton, S. Y. Savrasov, C. J. Humphreys, and A. P. Sutton, \textit{Electron-energy-loss spectra and the structural stability of nickel oxide: An LSDA+U study}, Phys. Rev. B \textbf{57}, 1505 (1998).

\bibitem{Darancet2014} P. Darancet, A. J. Millis, and C. A. Marianetti, \textit{Three-dimensional metallic and two-dimensional insulating behavior in octahedral tantalum dichalcogenides}, Phys. Rev. B \textbf{90}, 045134 (2014).

\bibitem{Spijkerman1997} A. Spijkerman, J. L. de Boer, A. Meetsma, and G. A. Wiegers, \textit{X-ray crystal-structure refinement of the nearly commensurate phase of 1T-TaS$_2$ in (3+2)-dimensional superspace}, Phys. Rev. B \textbf{56}, 13757 (1997).

\bibitem{Givens1977} F. Givens and G. Fredericks, \textit{Thermal expansion of NbSe$_{2}$ and TaS$_{2}$}, J. Phys. Chem. Solids \textbf{38}, 1363 (1977).


	

	


				


	


	




	





\end{thebibliography}

\begin{thebibliography}{99}

\bibitem{vonWitte2019} G. von Witte, T. Ki\ss linger, J. G. Horstmann, K. Rossnagel, M. A. Schneider, C. Ropers, and L. Hammer, \textit{Surface structure and stacking of the commensurate ($\sqrt{13}\times\sqrt{13}$)R13.9$^{\circ}$ charge density wave phase of 1T-TaS$_{2}$(0001)}, Phys. Rev. B \textbf{100}, 155407 (2019). 

\bibitem{Blum2001} V. Blum and K. Heinz, \textit{Fast LEED intensity calculations for surface crystallography using Tensor LEED}, Comput. Phys. Commun. \textbf{134}, 392 (2001).

\bibitem{Lee2019} S. -H. Lee, J. S. Goh, and D. Cho, \textit{Origin of the insulating phase and first-order metal-insulator transition in 1T-TaS$_{2}$}, Phys. Rev. Lett. \textbf{122}, 106404 (2019).

\bibitem{Kresse1996} G. Kresse and J. Furthm\"{u}ller, \textit{Efficient iterative schemes for ab initio total-energy calculations using a plane-wave basis set}, Phys. Rev. B \textbf{54}, 11169 (1996). 

\bibitem{Kresse1999} G. Kresse and D. Joubert, \textit{From ultrasoft pseudopotentials to the projector augmented-wave method}, Phys. Rev. B \textbf{59}, 1758 (1999). 

\bibitem{Blochl1994} P. E. Bl\"{o}chl, \textit{Projector augmented-wave method}, Phys. Rev. B \textbf{50}, 17953 (1994). 

\bibitem{Darancet2014} P. Darancet, A. J. Millis, and C. A. Marianetti, \textit{Three-dimensional metallic and two-dimensional insulating behavior in octahedral tantalum dichalcogenides}, Phys. Rev. B \textbf{90}, 045134 (2014).

\bibitem{Perdew1996} J. P. Perdew, K. Burke, and M. Ernzerhof, \textit{Generalized Gradient Approximation Made Simple}, Phys. Rev. Lett. \textbf{77}, 3865 (1996). 

\end{thebibliography}
